\documentclass{article}
\usepackage{graphicx} 
\usepackage{authblk}
\usepackage{mwe} 
\usepackage{blindtext} 
\usepackage{siunitx} 
\usepackage{mathtools} 
\usepackage{chemformula} 
\usepackage[utf8]{inputenc}
\usepackage[style=phys,sorting=none,backend=biber]{biblatex} 
\usepackage{amsmath,amssymb}
\usepackage{bm}
\usepackage{listings} 
\usepackage{graphicx}
\usepackage{lmodern}
\usepackage[textwidth=16cm,textheight=23cm]{geometry}

\usepackage{url}

\usepackage[font=small, labelfont=bf]{caption}
\usepackage{subfig} 
\usepackage{float}

\usepackage{booktabs}   
\usepackage{longtable}
\usepackage{array}
\usepackage{threeparttablex}
\usepackage[nameinlink, capitalise]{cleveref}

\usepackage{pgfplots}
\pgfplotsset{width=5cm,compat=1.9}

\usepgfplotslibrary{external}
\newcolumntype{x}[1]{%
>{\centering\hspace{0pt}}p{#1}}%

\usepackage{xr}
\makeatletter
\newcommand*{\addFileDependency}[1]{
  \typeout{(#1)}
  \@addtofilelist{#1}
  \IfFileExists{#1}{}{\typeout{No file #1.}}
}
\makeatother

\newcommand*{\myexternaldocument}[1]{%
    \externaldocument{#1}%
    \addFileDependency{#1.tex}%
    \addFileDependency{#1.aux}%
}
\myexternaldocument{SI}

\usepackage{enumitem}

\renewcommand{\vec}[1]{{\bf #1}} 

\usepackage{pstricks,pst-node,pst-tree} 
\usepackage{pst-pdf} 
\usepackage{pst-node}
\usepackage{pst-plot}
\usepackage{pst-3dplot}
\usepackage[b]{esvect}
\usepackage{chngcntr}

\definecolor{lbcolor}{rgb}{0.9,0.9,0.9}  

\author[1]{Juliane H. Fuglsbjerg\footnote{Email: jf@chem.ku.dk}}
\author[2]{Peter Reinholdt}
\author[2]{Erik Kjellgren}
\author[1]{Phillip W. K. Jensen}
\author[3]{Sonia Coriani}
\author[2]{Jacob Kongsted}
\author[1]{Stephan P. A. Sauer}
\affil[1]{\textit{Department of Chemistry, University of Copenhagen, Universitetsparken 5, DK-2100 Copenhagen \O, Denmark.}} \affil[2]{\textit{Department of Physics, Chemistry and Pharmacy,
University of Southern Denmark, Campusvej 55, DK-5230 Odense M, Denmark.}}
\affil[3]{\textit{Department of Chemistry, Technical University of Denmark, Kemitorvet Building 207, DK-2800 Kongens Lyngby, Denmark.}}
\date{}
\title{Orbital-Optimized Unitary Coupled Cluster for Indirect Nuclear Spin-Spin Coupling Constants within a Quantum Linear Response Framework}

\begin{document}
\maketitle
\begin{abstract} 
We present a quantum linear response (qLR) approach within an active-space framework for computing indirect nuclear spin-spin coupling constants, a key ingredient in NMR spectra predictions. The method employs the unitary coupled cluster (UCC) ansatz and its orbital-optimized variant (ooUCC), both suitable for quantum computing implementations, to evaluate spin-spin coupling constants via qLR. Test calculations on five small molecules are compared with CASCI, CASSCF, and conventional CCSD results. qLR with UCC/ooUCC yields spin-spin coupling constants comparable to classical methods. We further examine the role of orbital optimization and find that ooUCC markedly affects the computed couplings; orbital-optimized results show better agreement with CCSD. These findings indicate that orbital optimization is important for accurate NMR coupling predictions within quantum-computing-friendly correlated methods.
\end{abstract}

\section{Introduction}

Quantum computing provides a paradigm shift for simulating quantum systems by exploiting quantum superposition and entanglement. 
One prominent proposed application of quantum computing is within quantum chemistry, where, in particular, the solution of the electronic Schr{\"o}dinger equation is an important target \cite{cao_quantum_2019,bauer_quantum_2020,motta_emerging_2022}. 
In contrast to classical approaches, which require resources that grow exponentially with system size to represent the full wavefunction, quantum algorithms can, in principle, encode and manipulate these states efficiently. This capability makes quantum computing especially appealing for studying strongly correlated systems and large active spaces.
While most quantum algorithms developed to date have primarily targeted ground-state energies, there is growing interest in computing response properties—such as excitation energies, polarizabilities, and other molecular response properties \cite{cai2020quantum, Kumar2023, huang2022variational, Reinholdt2024,ziems_which_2024, jensen_quantum_2024, Von_Buchwald2024-pp}.

To the best of our knowledge, this work presents the first implementation of indirect nuclear spin-spin coupling constants within a quantum linear response framework. We employ quantum linear response (qLR) theory within an active space framework to calculate indirect nuclear spin-spin coupling constants, building upon the recently developed approaches of Ziems et al. \cite{ziems_which_2024} and Jensen et al. \cite{jensen_quantum_2024}. 

Indirect nuclear spin-spin coupling constants are a key property in predicting and interpreting the spectra of nuclear magnetic resonance (NMR) spectroscopy, which is a powerful tool for studying the structure of primarily closed-shell molecules. It relies on the interaction of the nuclear spins with an external magnetic field as well as the interaction between nuclear spins in a molecule. The latter is responsible for the splitting of peaks in NMR spectra. The indirect nuclear spin-spin coupling constant describes the indirect electron-coupled interaction between two nuclear spins in a molecule \cite{ramsey_electron_1953}.
Spin-spin coupling constants play an important role in the study of the stereochemistry of molecules \cite{Stereo1, Stereo2}, non-bonded interactions \cite{nonBond1, nonBond2} and tautomer equilibria \cite{Tauto}, to name a few. 
However, accurate calculations of spin-spin coupling constants require high-order coupled cluster wavefunction methods such as CC3 \cite{CC3a, CC3b, CC3c, CC3d}, which makes them challenging on classically based computers. Spin-spin coupling constants should therefore be a type of molecular property that can greatly benefit from the use of quantum computers for their accurate calculations.

The indirect nuclear spin-spin coupling constant consist of four contributions\cite{ramsey_electron_1953, spasB8}:
the diamagnetic spin-orbit contribution (DSO), the paramagnetic spin-orbit (PSO), the Fermi contact (FC), and the spin dipolar (SD) contributions.
The DSO term can be evaluated as a simple expectation value of the ground state wavefunction, although it can also be reformulated as a linear response function \cite{spas007, nmr12-jcp137-074108}. The remaining three terms are related to linear response functions, with the PSO term arising from an imaginary singlet perturbation, and the FC and SD terms from real triplet perturbations \cite{ramsey_electron_1953, spasB8}.
In this work we calculate indirect nuclear spin-spin coupling constants utilizing the variational quantum eigensolver and the unitary coupled cluster ansatz, as well as their orbital-optimized counterparts, within an active space approximation. In particular we extend the qLR method introduced by Ziems et al. \cite{ziems_which_2024} and Jensen et al. \cite{jensen_quantum_2024} to a triplet spin-adapted operator manifold necessary for the FC and SD contributions. 
The impact of including the orbital rotations in the wavefunction and qLR, as well as of truncating the active space, is investigated by computing the indirect nuclear spin-spin coupling constants for a selection of small molecules in a variety of active spaces and comparing them to the classical methods, including complete active space configuration interaction (CASCI), complete active space self consistent field (CASSCF) and conventional full-space coupled cluster singles doubles (CCSD).

\section{Theory}
In this section, we introduce the equations for calculating spin-spin coupling constants in a framework suitable for quantum computers. To do so, we first introduce indirect nuclear spin-spin coupling constants and their calculation with linear response theory. Then we briefly summarize the active space approximation and the unitary coupled cluster (UCC) and the orbital-optimized UCC (ooUCC) ans\"atze, followed by a review of quantum linear response theory in the active space framework.

Throughout this paper, the indices $p$, $q$, $r$, $s$ refer to general orbitals, $a$, $b$, $c$, $d$ refer to virtual orbitals, $i$, $j$, $k$,  $l$ refer to inactive orbitals, $v$, $w$, $x$, $z$ refer to active orbitals. Active orbitals with subscript refer to active orbitals that were, respectively, doubly occupied $v_i$ and virtual $v_a$ in the Hartree-Fock reference state.

\subsection{Nuclear Spin-Spin Coupling Constants}
The change in energy arising from the interaction of two nuclear moments $\vec{m}_A$ and $\vec{m}_B$ through the electrons may be expressed as
\begin{equation}
    \Delta E = \sum_{A > B}K^{AB}\vec{m}_A\vec{m}_B
\end{equation}
where $K^{AB}$ is the reduced indirect nuclear spin-spin coupling constant between nuclei $A$ and $B$. The reduced indirect spin-spin coupling constant is defined from the trace of a spin-spin coupling tensor
\begin{equation}
    K^{AB} = \frac{1}{3} \mathrm{Tr}(\vec{K}^{AB})
\end{equation}
The reduced indirect nuclear spin-spin coupling constant is independent of the nuclear g-factors $g_A$ and $g_B$. 
Because of this $K^{AB}$ cannot be measured from NMR spectra directly, but it can be related to the indirect nuclear spin-spin coupling constant, $J^{AB}$, which can
be measured
\begin{equation}
    K^{AB} = J^{AB} \frac{2\pi}{\mu_N^2 g_A g_B}~.
\end{equation}
where $\mu_N$ is the nuclear magneton. The nuclear magnetic moments of the nuclei, $\vec{m}^A$ and $\vec{m}^B$, are not included in the unperturbed Hamiltonian or wavefunction and thus need to be introduced. The wavefunction response to the perturbation can be found through the response function as discussed in \cref{sec:qLR}.
The first-order perturbation in the Hamiltonian consists of three contributions: the paramagnetic spin-orbit operator (PSO), the Fermi contact (FC) operator, and the spin dipolar (SD) operator. The second-order perturbation consists of the diamagnetic spin-orbit (DSO) operator. They are given as \cite{ramsey_electron_1953, spasB8}
\begin{equation}
    \hat{H}^{(1)} = - \sum_A \sum_\alpha \left( \hat{O}^\mathrm{PSO}_{A,\alpha} + \hat{O}^\mathrm{FC}_{A,\alpha} + \hat{O}^\mathrm{SD}_{A,\alpha} \right) m_{A,\alpha}
\end{equation} 
\begin{equation}
    \hat{H}^{(2)} = \sum_{AB} \sum_{\alpha\beta} \hat{O}^\mathrm{DSO}_{AB,\alpha\beta} m_{A,\alpha} m_{B,\beta}
\end{equation}
where $\alpha$ and $\beta$ are Cartesian components. The operators are defined as  \cite{vahtras_indirect_1992}
\begin{align}
    \hat{O}^\mathrm{PSO}_{A,\alpha} =& 
    - 
    \sum_{pq} 
    \left\langle p \left| 
    \frac{\hat{l}_\alpha(\mathbf{R}_A)}{|\mathbf{r} - \mathbf{R}_A|^3} 
    \right| q \right\rangle 
    \hat{E}_{pq} \label{eq:pso} \\
    \hat{O}^\mathrm{FC}_{A,\alpha} =& 
    -\frac{g_e4\pi}{3} 
    \sum_{pq} 
    \left\langle p \left| 
    \delta(\mathbf{r}-\mathbf{R}_K)
    \right| q \right\rangle 
    \hat{T}_{pq}^\alpha \label{eq:fc} \\
    \hat{O}^\mathrm{SD}_{A,\alpha} =& 
    -\frac{g_e}{2} 
    \sum_{pq} \sum_\beta 
    \left\langle p \left| 
    \frac{3 (r_\alpha - R_{A,\alpha})(r_\beta - R_{A,\beta}) - \delta_{\alpha\beta}|\mathbf{r}-\mathbf{R}_K|^2}{|\mathbf{r}-\mathbf{R}_K|^5} 
    \right| q \right\rangle 
    \hat{T}^\beta_{pq} \label{eq:sd} \\
    \hat{O}^\mathrm{DSO}_{AB,\alpha\beta} =& 
    \frac{1}{2} 
    \sum_{pq} 
    \left\langle p \left| 
    \frac{\delta_{\alpha\beta}
    (\mathbf{r}-\mathbf{R}_A)\cdot(\mathbf{r}-\mathbf{R}_B) 
    - (r_\alpha - R_{B,\alpha})(r_\beta -R_{A,\beta})}{|\mathbf{r}-\mathbf{R}_A|^3|\mathbf{r}-\mathbf{R}_B|^3} 
    \right| q \right\rangle 
    \hat{E}_{pq} \label{eq:dso}
\end{align}
where $g_e$ is the electronic g-factor, $p$ and $q$ are spatial orbitals, $\hat{l}$ is the angular momentum operator, $\mathbf{r}$ is the electron position vector, $\mathbf{R}_A$ is the position vector of nucleus $A$, $\hat{E}_{pq}$ is the singlet one-electron excitation operator from orbital $q$ to orbital $p$,
\begin{equation}
    \hat{E}_{pq} = \hat{a}^\dagger_{p\alpha}\hat{a}_{q\alpha} + \hat{a}^\dagger_{p\beta}\hat{a}_{q\beta}~,
\end{equation}
and $\hat{T}^{\alpha}_{pq}$ are Cartesian triplet one-electron excitation operators from orbital $q$ to orbital $p$,
\begin{align}
    \hat{T}^{x}_{pq} =& \frac{1}{2}(\hat{a}^\dagger_{p\alpha}\hat{a}_{q\beta} - \hat{a}^\dagger_{p\beta}\hat{a}_{q\alpha})~, \\
    \hat{T}^{y}_{pq} =& -\frac{1}{2i}(\hat{a}^\dagger_{p\alpha}\hat{a}_{q\beta} + \hat{a}^\dagger_{p\beta}\hat{a}_{q\alpha})~, \\
    \hat{T}^{z}_{pq} =& \frac{1}{2} (\hat{a}^\dagger_{p\alpha}\hat{a}_{q\alpha} -\hat{a}^\dagger_{p\beta}\hat{a}_{q\beta}) 
    ~.
\end{align}
The angular momentum operator is purely imaginary, and therefore so is the perturbation from the PSO operator. 
The remaining operators are real, and so are the corresponding perturbations. The FC and SD operators both operate on the spin of the wavefunction; When the ground state is a singlet, they require the wavefunction response to be a triplet in order to give non-zero contributions. The PSO operator does not operate on the spin and therefore requires the wavefunction response to have the same spin as the ground-state to give non-zero contributions.

The reduced indirect nuclear spin-spin coupling constant tensor can then be calculated as the derivative of the (quasi-)energy $E$ \cite{helgaker_recent_2012} with respect to the perturbation (nuclear magnetic moments) evaluated at zero perturbation,
\begin{equation}
\begin{split}
    K^{AB}_{\alpha\beta} =& \frac{\partial^2 E}{\partial m_{A,\alpha}\partial m_{B,\beta}} \Bigg|_{|\vec{m}_A|=|\vec{m}_B|=0}
    \\ =& \langle 0 | \hat{O}^\mathrm{DSO}_{AB,\alpha\beta} | 0 \rangle 
    + \langle\langle \hat{O}^\mathrm{PSO}_{A,\alpha} + \hat{O}^\mathrm{FC}_{A,\alpha} + \hat{O}^\mathrm{SD}_{A,\alpha} ;\hat{O}^\mathrm{PSO}_{B,\beta} + \hat{O}^\mathrm{FC}_{B,\beta} + \hat{O}^\mathrm{SD}_{B,\beta} \rangle\rangle 
    \\ =& \langle 0 | \hat{O}^\mathrm{DSO}_{AB,\alpha\beta} | 0 \rangle 
    + \langle\langle \hat{O}^\mathrm{PSO}_{A,\alpha};\hat{O}^\mathrm{PSO}_{B,\beta} \rangle\rangle 
    + \langle\langle \hat{O}^\mathrm{FC}_{A,\alpha};\hat{O}^\mathrm{FC}_{B,\beta} \rangle\rangle 
    + \langle\langle \hat{O}^\mathrm{SD}_{A,\alpha};\hat{O}^\mathrm{SD}_{B,\beta} \rangle\rangle + \langle \langle \hat{O}^\mathrm{FC}_{A,\alpha} ; \hat{O}^\mathrm{SD}_{B,\beta} \rangle\rangle + \langle\langle\hat{O}^\mathrm{SD}_{A,\alpha} ; \hat{O}^\mathrm{FC}_{B,\beta} \rangle \rangle~.
\end{split} 
\end{equation}
where $|0\rangle$ is the ground state wavefunction and $\langle\langle\hat{A};\hat{B}\rangle\rangle$ denotes the linear response function discussed in \cref{sec:qLR}.
There are no cross-terms between the PSO and the FC or SD operator because of the difference in spin. 
Note that only the trace of the spin-spin coupling tensor is needed for the indirect nuclear spin-spin coupling constant, which eliminates the FC/SD cross-terms as these are purely anisotropic. 

Both the FC \cref{eq:fc} and SD \cref{eq:sd} operators contain all three cartesian components of the triplet excitation operator which requires the wavefunction response and therefore the involved triplet spin-adapted excitation operators to allow for
all three orientations of the spin $M_S=-1,0,1$. However, the expression can be rewritten to a form that only contains the $z$-component of the triplet excitation operator, which only requires one orientation of the spin $M_S=0$ \cite{vahtras_indirect_1992}. This results in a simplified expression for the indirect nuclear spin-spin coupling constant.
\begin{equation} \label{eq:sscc}
    K^{AB}_{\alpha\alpha} = \langle 0 | \hat{O}^\mathrm{DSO}_{AB,\alpha\alpha} | 0 \rangle 
    + \langle\langle \hat{O}^\mathrm{PSO}_{A,\alpha};\hat{O}^\mathrm{PSO}_{B,\alpha} \rangle\rangle 
    + \langle\langle \hat{\Omega}^\mathrm{FC}_{A};\hat{\Omega}^\mathrm{FC}_{B} \rangle\rangle 
    + \sum_{\beta} \langle\langle \hat{\Omega}^\mathrm{SD}_{A,\alpha\beta};\hat{\Omega}^\mathrm{SD}_{B,\alpha\beta} \rangle\rangle
\end{equation}
where $\hat{\Omega}^\mathrm{FC}_A$ and $\hat{\Omega}^\mathrm{SD}_{A,\alpha\beta}$ are defined as
\begin{align}
    \hat{\Omega}^\mathrm{FC}_A =& 
    -\frac{g_e4\pi}{3} 
    \sum_{pq} 
    \left\langle p \left| 
    \delta(\mathbf{r}-\mathbf{R}_A)
    \right| q \right\rangle 
    \hat{T}_{pq}^z~, 
    \\
\hat{\Omega}^\mathrm{SD}_{A,\alpha\beta} =&
    -\frac{g_e}{2}
    \sum_{pq} 
    \left\langle p \left| 
    \frac{3 (r_\alpha - R_{A,\alpha})(r_\beta - R_{A,\beta}) - \delta_{\alpha\beta}|\mathbf{r}-\mathbf{R}_A|^2}{|\mathbf{r}-\mathbf{R}_A|^5} 
    \right| q \right\rangle 
    \hat{T}^z_{pq}~.
\end{align}

\subsection{Unitary coupled cluster within the active space approximation}
\label{wavefunction}

A well-known approximation in classical quantum chemistry is the active space approximation \cite{roos_complete_1980, siegbahn_comparison_1980, siegbahn_complete_1981}, where the wavefunction is partitioned into three subspaces: an inactive space $|I\rangle$ with doubly occupied orbitals, an active space $|A\rangle$ parametrized by some ansatz, $U(\boldsymbol{\theta})$, and a virtual space $|V\rangle$ where all orbitals are unoccupied.
\begin{equation}
    |\Psi(\boldsymbol{\theta})\rangle = |I\rangle \otimes U(\boldsymbol{\theta})|A\rangle \otimes |V\rangle = |I\rangle \otimes |A(\boldsymbol{\theta})\rangle \otimes |V\rangle
\end{equation}
Similarly, operators can be expressed in terms of contributions acting within each subspace, which yields expectation values in three parts.
\begin{equation}
    \hat{O} = \hat{O}_I \otimes \hat{O}_A \otimes \hat{O}_V
\end{equation}
\begin{equation}
    \langle \Psi(\boldsymbol{\theta}) | \hat{O} | \Psi(\boldsymbol{\theta}) \rangle = \langle I | \hat{O}_I | I \rangle \otimes \langle A(\boldsymbol{\theta}) | \hat{O}_A | A(\boldsymbol{\theta}) \rangle \otimes \langle V | \hat{O}_V | V \rangle
\end{equation}
Since only the active space is parameterized, it is the only space that needs to be simulated on quantum hardware \cite{mizukami_orbital_2020,sokolov_quantum_2020,Bierman2023-dm,ziems_which_2024}.

A popular chemistry-inspired ansatz for quantum computing is the unitary coupled cluster singles doubles (UCCSD) ansatz \cite{bartlett_alternative_1989,peruzzo_variational_2014}. 
It is defined as an exponentiated 
anti-Hermitian operator acting on a reference state, e.g., the Hartree-Fock state, in the full space $|\Phi_0\rangle$.
\begin{equation}
    |\mathrm{UCC}(\boldsymbol{\theta})\rangle = e^{\hat{T}_1(\boldsymbol{\theta}) - \hat{T}_1^\dagger(\boldsymbol{\theta}) + \hat{T}_2(\boldsymbol{\theta}) - \hat{T}_2(\boldsymbol{\theta})^\dagger} |\Phi_0 \rangle
\end{equation}
%
%
%
where 
$\hat{T}_1(\boldsymbol{\theta})$ and $\hat{T}_2(\boldsymbol{\theta})$ are singlet spin-adapted single and double excitation operators \cite{paldus_application_1977, piecuch_orthogonally_1989, packer_new_1996}, which within the active space approximation only acts on the active space.
\begin{equation}
\begin{split}
    \hat{T}_1(\boldsymbol{\theta}) =& \sum_{v_i v_a} \theta_{v_i}^{v_a} \frac{1}{\sqrt{2}} \hat{E}_{v_a v_i}
    \\ \hat{T}_2(\boldsymbol{\theta}) =& \sum_{\substack{v_i \geq v_j \\ v_a \geq v_b}} \theta_{v_i v_j}^{v_a v_b} \frac{1}{2}\frac{1}{\sqrt{(1+\delta_{v_a v_b})(1+\delta_{v_i v_j})}} (\hat{E}_{v_a v_i}\hat{E}_{v_b v_j} + \hat{E}_{v_a v_j}\hat{E}_{v_b v_i}) \\ & + \sum_{\substack{v_i > v_j \\ v_a > v_b}} \theta_{v_i v_j}^{v_a v_b} \frac{1}{2\sqrt{3}} (\hat{E}_{v_a v_i}\hat{E}_{v_b v_j} - \hat{E}_{v_a v_j}\hat{E}_{v_b v_i})~.
\end{split}
\end{equation}
The energy of the state is evaluated as an expectation value
\begin{equation}
    E(\boldsymbol{\theta}) = \langle \mathrm{UCC}(\boldsymbol{\theta}) | \hat{H} | \mathrm{UCC}(\boldsymbol{\theta}) \rangle
\end{equation}
with the Hamiltonian in second quantization, $\hat{H}$, given as
\begin{equation}
    \hat{H} = \sum_{pq} \hat{E}_{pq}h_{pq} + \frac{1}{2} \sum_{pqrs} \hat{e}_{pqrs} g_{pqrs}~,
\end{equation}
where $\hat{e}_{pqrs}$ is a two-body operator defined as
\begin{equation}
    \hat{e}_{pqrs} = \hat{E}_{pq}\hat{E}_{rs} - \delta_{qr} \hat{E}_{ps}
\end{equation}
and $h_{pq}$ and $g_{pqrs}$ are one- and two-electron integrals, respectively, in the molecular orbital (MO) basis. The wavefunction parameters, $\boldsymbol{\theta}$, are optimized by variational minimization. Using a quantum device to evaluate the energy is known as the variational quantum eigensolver (VQE) \cite{peruzzo_variational_2014,mcclean_theory_2016}.

The Hartree-Fock orbitals are, however, often not the best choice of orbitals. They can be improved by optimizing them such that the gradient of the energy with respect to the orbital rotation is zero. This results in an orbital-optimized UCCSD ansatz (ooUCCSD) \cite{mizukami_orbital_2020,Bierman2023-dm,sokolov_quantum_2020}, which further parametrizes the UCCSD wavefunction with an exponential orbital rotation operator
\begin{equation}
    |\mathrm{ooUCC}(\boldsymbol{\theta},\boldsymbol{\kappa})\rangle = e^{-\hat{\kappa}(\boldsymbol{\kappa})} |\mathrm{UCC}(\boldsymbol{\theta})\rangle~,
\end{equation}
where $\hat{\kappa}(\boldsymbol{\kappa})$ is defined as
\begin{equation}
    \hat{\kappa}(\boldsymbol{\kappa}) = \sum_{pq} \kappa^p_q \left( \hat{E}_{pq} - \hat{E}_{qp}\right) \qquad pq \in \{vi, ai, av\}.
\end{equation}
Only orbital rotations between spaces, i.e. inactive to active, inactive to virtual and active to virtual are included in the orbital rotation operator.

It is convenient to let the orbital rotation operator act on the Hamiltonian as this avoids explicit manipulation of the wavefunction. 
%
%
The energy expression for the ooUCC wavefunction is then
\begin{equation}
    E(\boldsymbol{\theta}, \boldsymbol{\kappa}) = \langle \mathrm{UCC}(\boldsymbol{\theta}) | \hat{H}(\boldsymbol{\kappa}) | \mathrm{UCC}(\boldsymbol{\theta}) \rangle ~,
\end{equation}
where
\begin{equation}
    \hat{H}(\boldsymbol{\kappa}) = e^{\hat{\boldsymbol{\kappa}}(\boldsymbol{\kappa})} \hat{H} e^{-\hat{\boldsymbol{\kappa}}(\boldsymbol{\kappa})}~.
\end{equation}
The parameters $\boldsymbol{\theta}$ and $\boldsymbol{\kappa}$ can then be found by variational minimization. Using a quantum device to evaluate the energy is known as orbital-optimized VQE (ooVQE) \cite{mizukami_orbital_2020,Bierman2023-dm,sokolov_quantum_2020}.

\subsection{Quantum Linear Response} \label{sec:qLR}
In this work, we use the qLR approach with active spaces introduced by Ziems et al. \cite{ziems_which_2024} and Jensen et al. \cite{jensen_quantum_2024}. 
%
%
%
%
Note that the groundstate wavefunction $|0\rangle$ in the following can be any variationally optimized wavefunction. In this work, we will use the UCC and ooUCC wavefunctions introduced in \cref{wavefunction}. 

The calculation of the indirect nuclear spin-spin coupling constant in \cref{eq:sscc} requires the linear response function \cite{olsen_linear_1985,helgaker_recent_2012} for two perturbation operators, $\hat{A}$ and $\hat{B}$, which is obtained as
\begin{equation} \label{eq:response_function}
    \langle \langle \hat{A} ; \hat{B} \rangle \rangle = -{\mathbf{V}_A^{[1]}}^\dagger \boldsymbol{\beta}_B
\end{equation}
where ${\mathbf{V}_A^{[1]}}^\dagger$ is the property gradient row vector and $\boldsymbol{\beta}_B$ is the linear response column vector. The latter is the solution to the linear response equation
\begin{equation} 
    \mathbf{E}^{[2]} \boldsymbol{\beta}_B = \mathbf{V}_B^{[1]}~,
\end{equation}
where $\mathbf{E}^{[2]}$ is the electronic Hessian. The electronic Hessian consists of the following sub-matrices
\begin{equation} \label{eq:hessian}
    \mathbf{E}^{[2]} = \begin{pmatrix}
        \mathbf{A} & \mathbf{B} \\ \mathbf{B}^* & \mathbf{A}^* 
    \end{pmatrix}
\end{equation}
\begin{equation} \label{eq:A_matrix}
    \mathbf{A} = \begin{pmatrix} \mathbf{A}^{qq} & \mathbf{A}^{qG} \\  \mathbf{A}^{Gq} & \mathbf{A}^{GG} \end{pmatrix} = \begin{pmatrix}
        \frac{1}{2} ( \langle  0 | [\hat{q}^\dagger_\mu, [\hat{H}, \hat{q}_{\mu '}]] | 0 \rangle + \mathrm{h.c.})
        & \frac{1}{2} ( \langle  0 | [\hat{G}_{n'}, [\hat{H}, \hat{q}^\dagger_{\mu}]] | 0 \rangle + \mathrm{h.c.})
        \\ \frac{1}{2} ( \langle  0 | [\hat{G}^\dagger_{n}, [\hat{H}, \hat{q}_{\mu '}]] | 0 \rangle + \mathrm{h.c.})
        & \frac{1}{2} ( \langle  0 | [\hat{G}^\dagger_n, [\hat{H}, \hat{G}_{n'}]] | 0 \rangle + \mathrm{h.c.})
    \end{pmatrix}
\end{equation}
\begin{equation} \label{eq:B_matrix}
    \mathbf{B} = \begin{pmatrix} \mathbf{B}^{qq} & \mathbf{B}^{qG} \\  \mathbf{B}^{Gq} & \mathbf{B}^{GG} \end{pmatrix} 
    = \begin{pmatrix}
        \frac{1}{2} ( \langle  0 | [\hat{q}^\dagger_\mu, [\hat{H}, \hat{q}^\dagger_{\mu '}]] | 0 \rangle + \mathrm{h.c.})
        & \frac{1}{2} ( \langle  0 | [\hat{G}^\dagger_{n'}, [\hat{H}, \hat{q}^\dagger_{\mu}]] | 0 \rangle + \mathrm{h.c.})
        \\ \frac{1}{2} ( \langle  0 | [\hat{G}^\dagger_{n}, [\hat{H}, \hat{q}^\dagger_{\mu '}]] | 0 \rangle + \mathrm{h.c.})
        & \frac{1}{2} ( \langle  0 | [\hat{G}^\dagger_n, [\hat{H}, \hat{G}^\dagger_{n'}]] | 0 \rangle + \mathrm{h.c.})~,
    \end{pmatrix}
\end{equation}
and property gradient is defined as
\begin{equation} \label{eq:PG_vector}
    \mathbf{V}_B^{[1]} = \begin{pmatrix}
        \mathbf{X}_B \\ -\mathbf{X}_B^*
    \end{pmatrix} = \begin{pmatrix}
        \langle 0 | [\hat{q}_\mu, \hat{B}] | 0 \rangle
        \\ \langle 0 | [\hat{G}_n, \hat{B}] | 0 \rangle
        \\ \langle 0 | [\hat{q}_\mu^\dagger, \hat{B}] | 0 \rangle
        \\ \langle 0 | [\hat{G}_n^\dagger, \hat{B}] | 0 \rangle
    \end{pmatrix}~.
\end{equation}
%
where $\hat{G}_n$ is an excitation operator within the active space and $\hat{q}_\mu$ is an orbital rotation operator between the spaces. Ziems et al. introduced multiple linear response parametrizations depending on the form of $\hat{G}_n$ and $\hat{q}_\mu$ \cite{ziems_which_2024}. In this work, the naive operators will be used, with the important distinction between the naive singlet spin-adapted operators, $\hat{q}^\mathrm{S}_\mu$ and $\hat{G}^\mathrm{S}_n$, and the naive triplet spin-adapted operators, $\hat{q}^\mathrm{T}_\mu$ and $\hat{G}^\mathrm{T}_n$. The excitation operators within the active space are truncated after the double excitations. The operators have the following form \cite{paldus_application_1977,oddershede_polarization_1984,piecuch_orthogonally_1989,packer_new_1996}
\begin{equation} \label{eq:q^S}
    \hat{q}^\mathrm{S}_\mu = \left\{ \frac{1}{\sqrt{2}} \hat{E}_{vi}, \frac{1}{\sqrt{2}} \hat{E}_{ai},\frac{1}{\sqrt{2}} \hat{E}_{av} \right\}~,
\end{equation}
\begin{equation} \label{eq:G^S}
    \hat{G}^\mathrm{S}_n = \left\{ \frac{1}{\sqrt{2}}\hat{E}_{v_a v_i}, \frac{1}{2 \sqrt{(1 + \delta_{v_av_b})(1+\delta_{v_iv_j})}} (\hat{E}_{v_av_i}\hat{E}_{v_bv_j} + \hat{E}_{v_av_j}\hat{E}_{v_bv_i}), \frac{1}{2\sqrt{3}}(\hat{E}_{v_av_i}\hat{E}_{v_bv_j} - \hat{E}_{v_av_j}\hat{E}_{v_bv_i}) \right\}~,
\end{equation}
\begin{equation} \label{eq:q^T}
    \hat{q}^\mathrm{T}_\mu = \left\{\frac{1}{\sqrt{2}} \hat{E}_{vi}^-,\frac{1}{\sqrt{2}} \hat{E}_{ai}^-,\frac{1}{\sqrt{2}} \hat{E}_{av}^-\right\}~,
\end{equation}
\begin{equation} \label{eq:G^T}
\begin{split}
    \hat{G}^\mathrm{T}_n = \left\{ 
    \frac{1}{\sqrt{2}}\hat{E}^-_{v_a v_i}, 
    \frac{(1-\delta_{v_iv_j})(1-\delta_{v_av_b})}{2 \sqrt{2}} (\hat{E}_{v_av_j}\hat{E}_{v_bv_i}^- + \hat{E}_{v_bv_i}\hat{E}_{v_av_j}^-), \right.
    \\ \frac{1}{2 \sqrt{(1-\delta_{v_iv_j})(1 + \delta_{v_av_b})}}( \hat{E}_{v_bv_j}\hat{E}_{v_av_i}^- + \hat{E}_{v_av_j}\hat{E}_{v_bv_i}^- ),
    \\ \left. \frac{1}{2\sqrt{(1+\delta_{v_iv_j})(1-\delta_{v_av_b})}}( \hat{E}_{v_bv_j}\hat{E}_{v_av_i}^- + \hat{E}_{v_bv_i}\hat{E}_{v_av_j}^- )
    \right\}~,
\end{split}
\end{equation}
where $\hat{E}_{pq}^-$ is the triplet spin-adapted one-electron excitation operator from orbital $q$ to orbital $p$ with $M_S=0$, defined as
\begin{equation}
    \hat{E}^{-}_{pq} = \hat{a}^\dagger_{p\alpha}\hat{a}_{q\alpha} - \hat{a}^\dagger_{p\beta}\hat{a}_{q\beta}~.
\end{equation}

\section{Computational details}
All calculations, except CCSD, were carried out using the active space approximation. The active spaces will be denoted ($n_e$,$n_o$), where $n_e$ is the number of electrons in the active space, and $n_o$ is the number of spatial orbitals in the active space. It is implied that the orbitals in the active space are the $n_e/2$ highest energy occupied and the $n_o-n_e/2$ lowest energy virtual orbitals and that all inactive orbitals are doubly occupied and all virtual orbitals are unoccupied. 

The molecular geometries were optimized at the Møller-Plesset second-order perturbation theory (MP2) level \cite{moller_note_1934} in a 6-31G basis set \cite{ditchfield_selfconsistent_1971,hehre_selfconsistent_1972} using the Gaussian program \cite{gausian03}.

The basis set employed in the calculation of the indirect nuclear spin-spin coupling constants is the 6-31G-J basis set \cite{kjaer_pople_2011} from basis set exchange \cite{schuchardt_basis_2007,pritchard_new_2019}, which is optimized for the calculations of indirect nuclear spin-spin coupling constants.

The UCCSD and ooUCCSD wavefunction and linear response calculations were carried out utilizing our in-house quantum chemistry software SlowQuant \cite{kjellgren_slowquant_2025} which is interfaced with PySCF \cite{sun_libcint_2015,sun_recent_2020} for the Hamiltonian one- and two-electron integrals, the initial Hartree-Fock orbitals as well as the PSO integrals and the sum of the FC and SD integrals. For ooUCCSD(6,5) on water, PySCF was also utilized to calculate initial CASSCF orbitals for the ooUCCSD calculation, since it was otherwise struggling with converging to the ground state.
CASSCF and CASCI indirect nuclear spin-spin coupling constant calculations were performed using the Dalton program \cite{aidas_dalton_2014, noauthor_dalton_2022}. Full-space CCSD indirect nuclear spin-spin coupling constant calculations were carried out using the CFOUR program \cite{matthews_coupled-cluster_2020}.
The CASSCF, CASCI, and CCSD results serve as reference values against which the UCC methods are benchmarked. These are chosen as reference values as there are too many missing effects (small basis set, solvent effect, vibrational correction) for comparison to experimental results to provide meaningful insights.

\section{Results and discussion}

\subsection{Hydrogen molecule}

One of the key differences between the quantum methods, UCCSD and ooUCCSD, and the classical methods, CASCI and CASSCF, is that 
in the quantum methods
the excitations within the active space are truncated after double excitations in both the wavefunction and the linear response function, whereas the classical methods include all possible excitations within the active space. However, for H$_2$, there are only two electrons, making double excitations the highest excitation order, and the methods are therefore expected to give identical results. In the 6-31G-J basis, (2,8) is the full space for H$_2$ so those solutions will be full CI (FCI).

\begin{figure}[h!]
    \centering
    \includegraphics[width=0.5\linewidth]{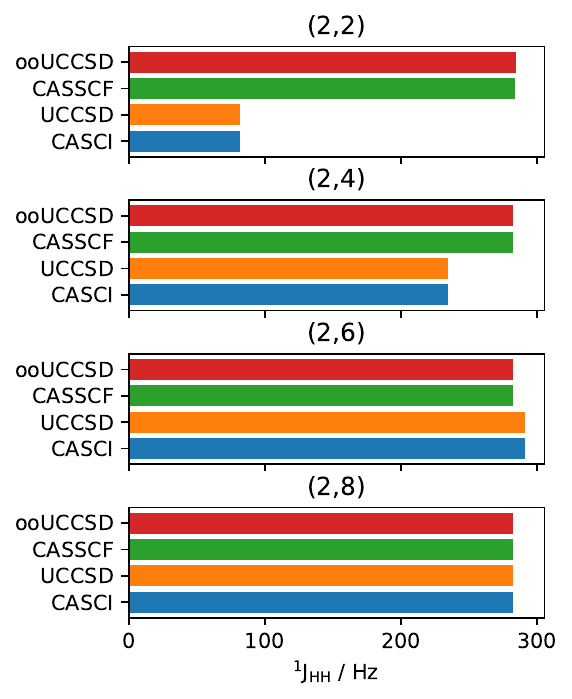}
    \caption{Spin-spin coupling constant of H$_2$ in a 6-31G-J basis for four different active-space methods and four different choices of active space.}
    \label{fig:h2_sscc}
\end{figure}

The one-bond coupling constant between the two hydrogens $^1J_\mathrm{HH}$ across four different active spaces can be seen in \cref{fig:h2_sscc} and in \cref{tab:h2_hh} in the Supporting Information.
The results for the methods without orbital optimization, UCCSD and CASCI, are indeed identical, as are the results for the orbital-optimized methods, ooUCCSD and CASSCF. 
The results of the orbital-optimized methods are nearly invariant with respect to the active space for H$_2$. For these methods the difference in the coupling constant is only $1.94\ \mathrm{Hz}$ between the smallest active space (2,2) and the full space (2,8), for which the results is $282.22\ \mathrm{Hz}$, i.e. only a 0.7\% difference. On the other hand, for the methods without orbital optimization, the difference in coupling constant between the smallest and largest active space is huge, i.e. $200.24\ \mathrm{Hz}$ or a 71.0\% difference. With increasing active spaces, the results without orbital optimization approach the FCI results but not in a monotonic way.

The breakdown of the $^1J_\mathrm{HH}$ coupling constant into its four contributions: DSO, PSO, FC and SD can be seen in \cref{fig:h2_hh_terms} in the Supporting Information. This shows that the $^1J_\mathrm{HH}$ coupling constant is entirely dominated by the FC contribution, as the DSO and SD contributions are smaller than 1 Hz and the PSO is zero for all active spaces and methods. 
Furthermore, the DSO and SD contributions of UCCSD and CASCI reach the FCI solution already for the (2,6) active space, whereas the FC term requires the full (2,8) space to reach the FCI solution.

\subsection{Water}

\begin{figure}[h!]
    \centering
    \subfloat[HH coupling]{\includegraphics[width=0.5\linewidth]{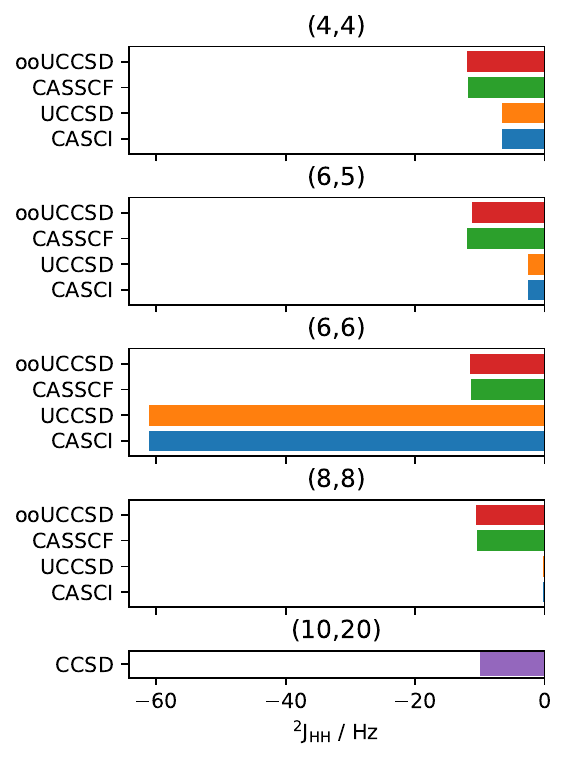}\label{fig:h2o_hh_sscc}}
    \subfloat[OH coupling]{\includegraphics[width=0.5\linewidth]{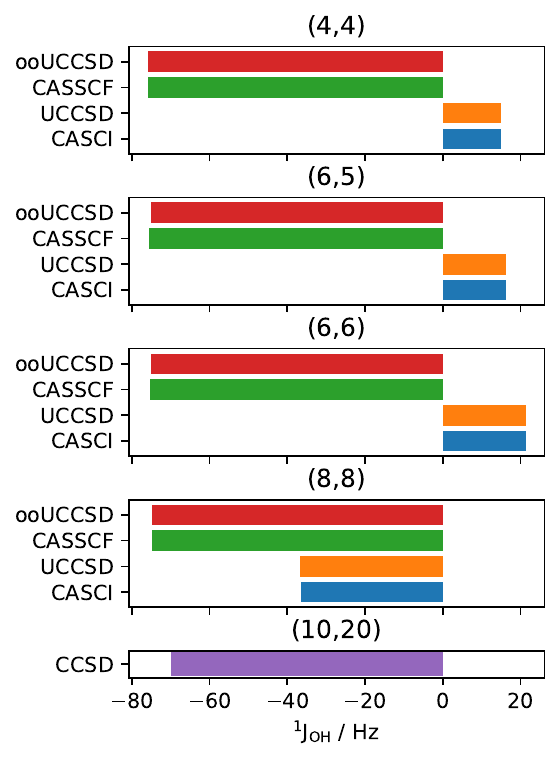}\label{fig:h2o_oh_sscc}}
    \caption{Spin-spin coupling constants of H$_2$O in a 6-31G-J basis for four different 
    active-space methods and four different choices of active space
    as well as for full-space (10,20) CCSD.}
    \label{fig:h2o_sscc}
\end{figure}

In water, there are two different coupling constants, a two-bond coupling between the two hydrogen atoms, $^2J_{\mathrm{HH}}$, and a one-bond coupling between one hydrogen atom and the oxygen atom, $^1J_{\mathrm{OH}}$. The results for both couplings across four active spaces and full-space CCSD are depicted in \cref{fig:h2o_sscc} and in \cref{tab:h2o_hh,tab:h2o_oh} in the Supporting Information. 

Examining the results of the two orbital-optimized methods, CASSCF and ooUCCSD, shows that these two methods yield very similar results. The largest deviation between the two methods is $0.79\ \mathrm{Hz}$ (or 6.6\%) for $^2J_{\mathrm{HH}}$ and $0.75\ \mathrm{Hz}$ 
(or 1.0\%) for $^1J_{\mathrm{OH}}$ in the (6,5) active space. 
This indicates that truncating after the double excitations in ooUCCSD is sufficient for this system. 
The ooUCCSD spin-spin coupling constants vary slightly but not significantly with the active space. The difference between the smallest and largest active space employed here is $1.41\ \mathrm{Hz}$ (or 13.3\%) for $^2J_{\mathrm{HH}}$ and $1.00\ \mathrm{Hz}$ (or 1.3\%) for $^1J_{\mathrm{OH}}$.

Comparing the results of the two methods without orbital optimization, CASCI and UCCSD, to each other shows that these results are close to each other with the largest deviation being $0.12\ \mathrm{Hz}$ (or 0.3\%) for the $^1J_{\mathrm{OH}}$ coupling in the (8,8) active space. This indicates that truncating after the double excitations is also sufficient for calculations without orbital optimization for this system. 
However, the coupling constants are highly dependent on the size of the active space in the non-orbital-optimized calculations.
The smaller active spaces even predict the wrong sign for the $^1J_{\mathrm{OH}}$ coupling constant.
For UCCSD the difference between the smallest active space (4,4) and the largest active space (8,8) is for $^2J_{\mathrm{HH}}$ $6.28\ \mathrm{Hz}$ and for $^1J_{\mathrm{OH}}$ $51.67\ \mathrm{Hz}$. However, the difference between the (6,6) and (8,8) active space is even larger for both couplings, so there is no sign of convergence with respect to the active space in the non-orbital-optimized calculations. 

The breakdown of the coupling constants into its individual contributions can be seen in \cref{fig:oh2_hh_terms,fig:oh2_oh_terms} in the Supporting Information. For UCCSD the DSO, PSO and SD contributions are all largely converged with respect to the active space, where the change from (6,6) to (8,8) is less than $0.15\ \mathrm{Hz}$ across both couplings. The FC contribution, however, is not converged with a change of about $60\ \mathrm{Hz}$ from (6,6) to (8,8) for both coupling constants. The FC term is also responsible for 
$^2J_{\mathrm{HH}}$ varying enormously from the (6,6) active space to the remaining three active spaces and for the change in sign of $^1J_{\mathrm{OH}}$ going from the (6,6) to the (8,8) active space.

Lastly, comparing the results of the methods without orbital optimization, CASCI and UCCSD, to their orbital-optimized counterparts, CASSCF and ooUCCSD, and with the results of the full-space CCSD calculations, it is evident that orbital optimization has a huge effect on the spin-spin coupling constants. Comparing results from the largest active space (8,8) to the full-space CCSD solution, the results of the orbital-optimized methods are in significantly better agreement with CCSD. 
The difference between the UCCSD and CCSD results is as much as $9.74\ \mathrm{Hz}$ (or 97.3\%) for $^2J_{\mathrm{HH}}$ and $33.36\ \mathrm{Hz}$ (or 47.7\%) for $^1J_{\mathrm{OH}}$ while the difference between the ooUCCSD and CCSD results is only $0.59\ \mathrm{Hz}$ (or 5.9\%) for $^2J_{\mathrm{HH}}$ and $4.76\ \mathrm{Hz}$ (or 6.8\%) for $^1J_{\mathrm{OH}}$.

\subsection{Ammonia}

\begin{figure}[h!]
    \centering
    \subfloat[HH coupling]{\includegraphics[width=0.5\linewidth]{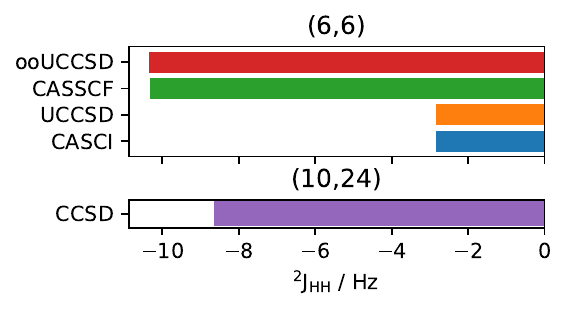}\label{fig:nh3_hh_sscc}}
    \subfloat[NH coupling]{\includegraphics[width=0.5\linewidth]{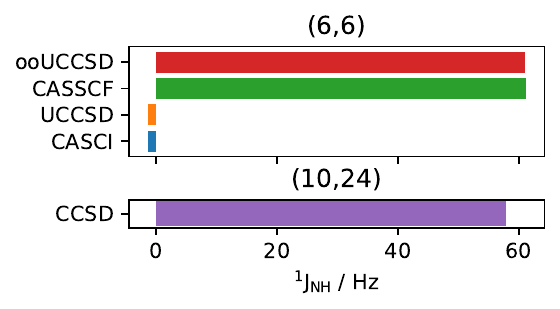}\label{fig:nh3_nh_sscc}}
    \caption{Spin-spin coupling constants of NH$_3$ in a 6-31G-J basis with four different active-space methods in a (6,6) active space as well as for full-space (10,24) CCSD.}
    \label{fig:nh3_sscc}
\end{figure}

The results of the two-bond hydrogen-hydrogen coupling constant $^2J_{\mathrm{HH}}$ and the one-bond hydrogen-nitrogen coupling constant $^1J_{\mathrm{NH}}$ in ammonia can be seen in \cref{fig:nh3_sscc}, and in \cref{tab:nh3_hh,tab:nh3_nh} in the Supporting Information for a (6,6) active space and 
full-space CCSD.

Comparing UCCSD to its classical counterpart CASCI, we find that the results are nearly identical. The largest deviation between the two methods is $0.01\ \mathrm{Hz}$ (or $0.6\%$) for the $^1J_\mathrm{NH}$ coupling. Similarly, the ooUCCSD and CASSCF results are close to each other, with the largest deviation being $0.10\ \mathrm{Hz}$ (or $0.2\%$) also for the $^1J_\mathrm{NH}$ coupling. This shows that also for ammonia is truncating the cluster expansion after double excitations in UCC is sufficient for the coupling constants.

Comparison of the results between the methods with and without orbital optimization reveals that the inclusion of orbital optimization drastically changes the resulting spin-spin coupling constants. Comparison to full-space CCSD results reveal that ooUCCSD is largely in agreement with CCSD while UCCSD is not.
For $^1J_\mathrm{NH}$, which deviates the most for both methods, ooUCCSD differs from CCSD by $3.16\ \mathrm{Hz}$ (or 5.5\%) while UCCSD differs by $59.25\ \mathrm{Hz}$ (or 102.4\%).
In case of the one-bond nitrogen-hydrogen coupling constant, UCCSD and CASCI even predict the wrong sign for the coupling constant in the (6,6) active space, similar to the results for the one-bond coupling in water.

In \cref{fig:nh3_hh_terms,fig:nh3_nh_terms} in the Supporting Information the individual contributions to the spin-spin coupling constants in ammonia are depicted. The primary contribution to the deviation between the UCC methods and CCSD is the FC contribution. For UCCSD the FC term of both coupling constants in ammonia has the opposite sign of the CCSD FC term. For ooUCCSD, the FC has the correct sign, but there is a slight difference in magnitude most noticeable for the $^1J_\mathrm{NH}$ coupling with a difference of $3.13\ \mathrm{Hz}$ (or $5.6\%$).

\subsection{Methane}

\begin{figure}[h!]
    \centering
    \subfloat[HH coupling]{\includegraphics[width=0.5\linewidth]{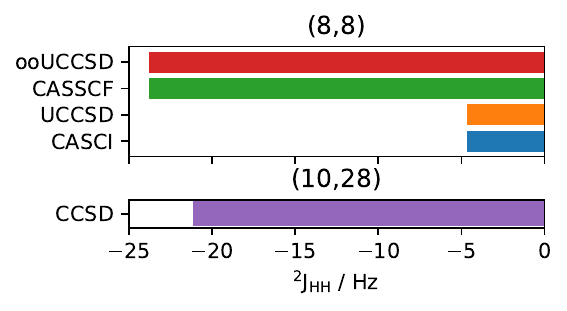}\label{fig:ch4_hh_sscc}}
    \subfloat[CH coupling]{\includegraphics[width=0.5\linewidth]{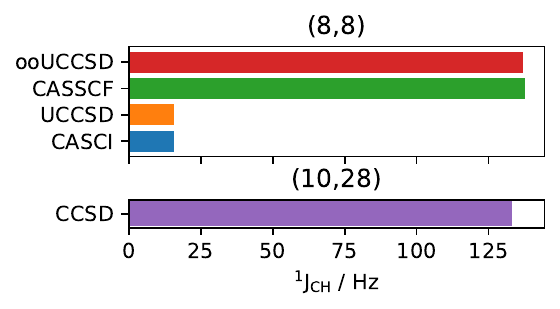}\label{fig:ch4_ch_sscc}}
    \caption{Spin-spin coupling constants of CH$_4$ in a 6-31G-J basis with four different active-space methods in an (8,8) active space and for full-space (10,28) CCSD.}
    \label{fig:ch4_sscc}
\end{figure}

The results for the two-bond hydrogen-hydrogen coupling constant $^2J_{\mathrm{HH}}$ and the one-bond hydrogen-carbon coupling constant $^1J_{\mathrm{CH}}$ in methane are depicted in \cref{fig:ch4_sscc}, and in \cref{tab:ch4_hh,tab:ch4_ch} in the Supporting Information for the four different active-space methods in an (8,8) active space, as well as full-space CCSD.

Comparing UCCSD to its classical counterpart CASCI, the methods yield very similar results with the largest deviation being $0.05\ \mathrm{Hz}$ 
(or $0.3\%$) for the $^1J_\mathrm{CH}$ coupling. Similarly, the results of ooUCCSD and CASSCF are close; The largest deviation between the two methods is $0.53\ \mathrm{Hz}$ 
(or $0.4\%$), which is also for the $^1J_\mathrm{CH}$ coupling. This shows that also for this calculation is truncating the cluster expansion after double excitations is sufficient.

The results vary drastically between UCCSD with and without orbital optimization. The UCCSD results are very different from the full-space CCSD results with a difference of $117.50\ \mathrm{Hz}$ 
(or 88.2\%) between the full-space CCSD and UCCSD(8,8) results for $^1J_\mathrm{CH}$. The ooUCCSD(8,8) results, however, are close to the full-space CCSD results with a deviation of $3.85\ \mathrm{Hz}$ 
(or 2.9\%) for the same coupling constant.

The individual contributions of the spin-spin coupling constants in methane are pictured in \cref{fig:ch4_hh_terms,fig:ch4_ch_terms} in the Supporting Information. Of the individual terms only DSO is similar across all five methods. The FC term is the primary source of deviation between CCSD and the UCC methods. ooUCCSD deviates with $3.92\ \mathrm{Hz}$ (or 3.0\%) for $^1J_\mathrm{CH}$ and UCCSD deviates with $115.83\ \mathrm{Hz}$ (or 88.2\%). For the other terms, UCCSD and ooUCCSD do not necessarily capture the correct behavior, but the terms are overall very small and so the effects are negligible compared to the FC terms in the final spin-spin coupling constants.

\subsection{Carbon monoxide}

\begin{figure}[h!]
    \centering
    \includegraphics[width=0.5\linewidth]{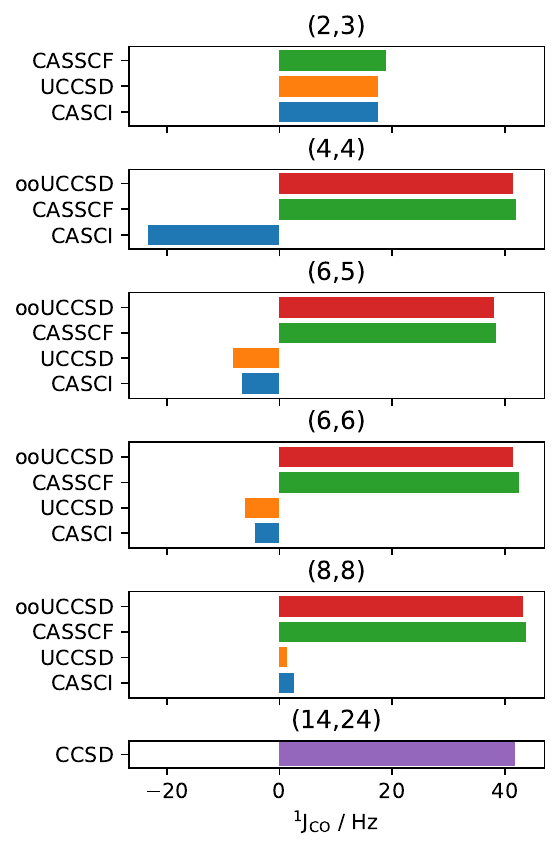}
    \caption{Spin-spin coupling constants of CO in a 6-31G-J basis for five different active spaces and four different active-space methods, as well as for full-space (14,24) CCSD.}
    \label{fig:co_sscc}
\end{figure}

The calculated one-bond carbon-oxygen $^1J_\mathrm{CO}$ coupling constant in carbon monoxide is depicted in \cref{fig:co_sscc},  and in \cref{tab:co_co} in the Supporting Information, across five different active spaces and full-space CCSD. The (2,3) active space consists of the HOMO, a $\sigma$ orbital, and the LUMOs, doubly degenerate $\pi^*$ orbitals. When including orbital rotations, the HOMO changes from the $\sigma$ orbital to doubly degenerate $\pi$ orbitals, which are then included instead to give a (4,4) active space. To avoid breaking the symmetry when selecting an active space, the CASSCF(2,3) and the CASCI(4,4) calculations have been carried out with the $\sigma$ and $\pi$ orbital(s) respectively, even though there is a higher lying occupied orbital which is not included in the active space. 
Handpicking orbitals is not built into our UCC implementation, therefore ooUCCSD(2,3) and UCCSD(4,4) have been excluded.

First, comparing UCCSD and CASCI, the two methods are still generally in agreement, but the deviations are larger now with the largest deviation between the two methods being $1.75\ \mathrm{Hz}$ or $26.7\%$ in the (6,5) active space. This deviation is significantly larger than the deviations in the previous systems for UCCSD and CASCI. The results are again very dependent on the size of the active space with the total spin-spin coupling constant going from a positive value for (2,3) to negative values for (4,4) to (6,6) and then back to a positive value for (8,8).

Comparing the results of the two orbital-optimized methods, ooUCCSD and CASSCF, the largest deviation between the two methods in CO is $0.99\ \mathrm{Hz}$ 
(or $2.3\%$) in the (6,6) active space. This is larger than but still comparable to the deviations in any of the previous systems. With the exception of the smallest active space (2,3), the total spin-spin coupling constant is relatively stable with respect to the size of the active space for ooUCCSD. However, investigating the individual terms in the $^1J_\mathrm{CO}$, depicted in \cref{fig:co_co_terms} in the Supporting Information, the PSO term decreases with the size of the active space while the FC term increases with the size of the active space and the stability in the total coupling constant is due to error cancellation, indicating that the smaller active spaces are not larger enough to capture the correct behavior of the $^1J_\mathrm{CO}$ coupling.

The inclusion of orbital optimization has a large effect on the resulting indirect nuclear spin-spin coupling constant. UCCSD(8,8) is in poor agreement with the full-space CCSD results deviating with $40.34\ \mathrm{Hz}$ 
(or $96.7\%$), while the ooUCCSD(8,8) results are in good agreement deviating with just $1.54\ \mathrm{Hz}$ 
(or $3.7\%$).
From the individual terms in \cref{fig:co_co_terms} in the Supporting Information, it can be seen that the poor agreement of UCCSD with CCSD is primarily due to drastic differences in the PSO and FC contributions.

\section{Conclusion}
In this paper we have reported the implementation of indirect nuclear spin-spin coupling constants within a quantum-computing-ready framework in the SlowQuant package---to our knowledge the first of such implementations. Our approach combines an active space quantum linear response (qLR) formalism using both singlet- and triplet-adapted operator manifolds with variationally optimized ground-state wavefunctions. 
In this study we employ the unitary coupled cluster (UCC) and orbital-optimized UCC (ooUCC) ans{\"a}tze and compute one- and two-bond couplings for H$_2$, H$_2$O, NH$_3$, CH$_4$, and CO.

Truncation of the active-space excitation operator at doubles (UCCSD/ooUCCSD) reproduces CASCI/CASSCF results for the systems studied, indicating that higher excitations within the active space are largely insignificant except for CO, where their effect is largest. Orbital optimization markedly improves robustness: ooUCCSD results are relatively insensitive to the active space choice and qualitatively agree with full-space CCSD even for small active spaces, whereas UCCSD shows strong active space dependence and fails to approach converged behavior in the spaces tested. 
Analysis of the individual contributions to the spin-spin coupling constants shows that the Fermi contact term--also the largest--is the most challenging to capture; orbital rotations improve the paramagnetic spin-orbit, spin-dipolar, and Fermi contact terms, while the diamagnetic spin-orbit term is the only contribution well described without orbital optimization.

These findings demonstrate the importance of orbital rotations for accurate NMR coupling predictions in quantum-computing-compatible correlated methods. Future work will assess the impact of probabilistic and simulated noise on the computed coupling constants.

\section*{Acknowledgements}
We acknowledge financial Support from the Novo Nordisk Foundation (NNF) for
the focused research project ``Hybrid Quantum Chemistry on
Hybrid Quantum Computers'' (NNF grant NNFSA220080996).

\printbibliography \pagebreak

\appendix
\counterwithin{figure}{section}
\counterwithin{table}{section}

\section{Tabulated spin-spin coupling constant results}

\begin{table}[h!]
    \caption{Two-bond hydrogen-hydrogen indirect nuclear spin-spin coupling constant $^2J_\mathrm{HH}$ in H$_2$O for CASCI, UCCSD, CASSCF and ooUCCSD in four different active spaces.} \label{tab:h2_hh}
    \centering 
    \begin{tabular}{rlccccc}
        \toprule \multicolumn{2}{l}{Method} & DSO & PSO & SD & FC & Total \\ \midrule
(2,2) & CASCI & $-0.39$ & $0.00$ & $0.40$ & $81.97$ & $81.98$ \\ & UCCSD & $-0.39$ & $0.00$ & $0.40$ & $81.97$ & $81.98$ \\ & CASSCF & $-0.29$ & $0.00$ & $0.71$ & $283.75$ & $284.16$ \\ & ooUCCSD & $-0.29$ & $0.00$ & $0.71$ & $283.75$ & $284.16$ \\ [1.0ex]
(2,4) & CASCI & $-0.25$ & $0.00$ & $0.88$ & $233.74$ & $234.36$ \\ & UCCSD & $-0.25$ & $0.00$ & $0.88$ & $233.74$ & $234.36$ \\ & CASSCF & $-0.29$ & $0.00$ & $0.72$ & $281.97$ & $282.40$ \\ & ooUCCSD & $-0.29$ & $0.00$ & $0.72$ & $281.97$ & $282.41$ \\ [1.0ex]
(2,6) & CASCI & $-0.29$ & $0.00$ & $0.72$ & $290.60$ & $291.03$ \\ & UCCSD & $-0.29$ & $0.00$ & $0.72$ & $290.60$ & $291.03$ \\ & CASSCF & $-0.29$ & $0.00$ & $0.72$ & $281.79$ & $282.22$ \\ & ooUCCSD & $-0.29$ & $0.00$ & $0.72$ & $281.79$ & $282.22$ \\ [1.0ex]
(2,8) & CASCI & $-0.29$ & $0.00$ & $0.72$ & $281.79$ & $282.22$ \\ & UCCSD & $-0.29$ & $0.00$ & $0.72$ & $281.79$ & $282.22$ \\ & CASSCF & $-0.29$ & $0.00$ & $0.72$ & $281.79$ & $282.22$ \\ & ooUCCSD & $-0.29$ & $0.00$ & $0.72$ & $281.79$ & $282.22$ \\ [1.0ex]
    \bottomrule \end{tabular}
\end{table}

\begin{table}[h!]
    \caption{Two-bond hydrogen-hydrogen indirect nuclear spin-spin coupling constant $^2J_\mathrm{HH}$ in H$_2$O for CASCI, UCCSD, CASSCF and ooUCCSD in four different active spaces and full-space (10,20) CCSD with the 6-31G-J basis set.} \label{tab:h2o_hh}
    \centering
    \begin{tabular}{rlccccc}
        \toprule \multicolumn{2}{l}{Method} & DSO & PSO & SD & FC & Total \\ \midrule
(4,4) & CASCI & $-6.92$ & $2.16$ & $0.01$ & $-1.81$ & $-6.57$ \\ & UCCSD & $-6.92$ & $2.16$ & $0.01$ & $-1.81$ & $-6.57$ \\ & CASSCF & $-6.86$ & $3.40$ & $1.03$ & $-9.49$ & $-11.91$ \\ & ooUCCSD & $-6.86$ & $3.41$ & $1.03$ & $-9.60$ & $-12.02$ \\ [1.0ex]
(6,5) & CASCI & $-6.88$ & $1.92$ & $0.28$ & $2.09$ & $-2.59$ \\ & UCCSD & $-6.89$ & $1.91$ & $0.28$ & $2.10$ & $-2.59$ \\ & CASSCF & $-6.85$ & $3.39$ & $1.03$ & $-9.51$ & $-11.94$ \\ & ooUCCSD & $-6.85$ & $3.40$ & $1.03$ & $-8.73$ & $-11.15$ \\ [1.0ex]
(6,6) & CASCI & $-6.87$ & $2.26$ & $0.24$ & $-56.76$ & $-61.13$ \\ & UCCSD & $-6.87$ & $2.26$ & $0.24$ & $-56.80$ & $-61.18$ \\ & CASSCF & $-6.86$ & $3.38$ & $1.00$ & $-8.91$ & $-11.39$ \\ & ooUCCSD & $-6.86$ & $3.38$ & $1.00$ & $-9.04$ & $-11.52$ \\ [1.0ex]
(8,8) & CASCI & $-6.83$ & $2.24$ & $0.20$ & $4.09$ & $-0.31$ \\ & UCCSD & $-6.83$ & $2.22$ & $0.20$ & $4.13$ & $-0.28$ \\ & CASSCF & $-6.87$ & $3.35$ & $0.97$ & $-7.95$ & $-10.50$ \\ & ooUCCSD & $-6.87$ & $3.35$ & $0.97$ & $-8.06$ & $-10.61$ \\ [1.0ex]
(10,20) & CCSD & $-6.94$ & $3.45$ & $0.95$ & $-7.48$ & $-10.02$ \\[1.0ex]
    \bottomrule \end{tabular}
\end{table}

\begin{table}[h!]
    \caption{One-bond oxygen-hydrogen indirect nuclear spin-spin coupling constant $^1J_\mathrm{OH}$ in H$_2$O for CASCI, UCCSD, CASSCF and ooUCCSD in four different active spaces and full-space (10,20) CCSD with the 6-31G-J basis set.} \label{tab:h2o_oh}
    \centering
    \begin{tabular}{rlccccc}
        \toprule \multicolumn{2}{l}{Method} & DSO & PSO & SD & FC & Total \\ \midrule
(4,4) & CASCI & $-0.31$ & $-2.62$ & $-0.61$ & $18.57$ & $15.03$ \\ & UCCSD & $-0.31$ & $-2.62$ & $-0.61$ & $18.57$ & $15.03$ \\ & CASSCF & $-0.32$ & $-10.35$ & $0.29$ & $-65.55$ & $-75.93$ \\ & ooUCCSD & $-0.32$ & $-10.32$ & $0.29$ & $-65.40$ & $-75.76$ \\ [1.0ex]
(6,5) & CASCI & $-0.32$ & $-3.67$ & $-0.10$ & $20.38$ & $16.29$ \\ & UCCSD & $-0.32$ & $-3.67$ & $-0.09$ & $20.43$ & $16.34$ \\ & CASSCF & $-0.32$ & $-10.33$ & $0.29$ & $-65.34$ & $-75.70$ \\ & ooUCCSD & $-0.32$ & $-10.26$ & $0.28$ & $-64.64$ & $-74.95$ \\ [1.0ex]
(6,6) & CASCI & $-0.32$ & $-3.52$ & $-0.04$ & $25.20$ & $21.32$ \\ & UCCSD & $-0.32$ & $-3.52$ & $-0.04$ & $25.24$ & $21.36$ \\ & CASSCF & $-0.31$ & $-10.08$ & $0.43$ & $-65.27$ & $-75.23$ \\ & ooUCCSD & $-0.31$ & $-10.04$ & $0.43$ & $-65.08$ & $-75.01$ \\ [1.0ex]
(8,8) & CASCI & $-0.33$ & $-3.66$ & $0.05$ & $-32.57$ & $-36.52$ \\ & UCCSD & $-0.33$ & $-3.66$ & $0.06$ & $-32.71$ & $-36.64$ \\ & CASSCF & $-0.31$ & $-9.89$ & $0.44$ & $-65.16$ & $-74.92$ \\ & ooUCCSD & $-0.31$ & $-9.85$ & $0.44$ & $-65.05$ & $-74.76$ \\ [1.0ex]
(10,20) & CCSD & $-0.31$ & $-10.00$ & $0.31$ & $-60.01$ & $-70.00$ \\[1.0ex]
    \bottomrule \end{tabular}
\end{table}

\begin{table}[h!]
    \caption{Two-bond hydrogen-hydrogen indirect nuclear spin-spin coupling constant $^2J_\mathrm{HH}$ in NH$_3$ for CASCI, UCCSD, CASSCF and ooUCCSD in one active space and full-space (10,24) CCSD with the 6-31G-J basis set.} \label{tab:nh3_hh}
    \centering
    \begin{tabular}{rlccccc}
        \toprule \multicolumn{2}{l}{Method} & DSO & PSO & SD & FC & Total \\ \midrule
(6,6) & CASCI & $-5.75$ & $1.08$ & $0.17$ & $1.66$ & $-2.83$ \\ & UCCSD & $-5.75$ & $1.08$ & $0.17$ & $1.66$ & $-2.84$ \\ & CASSCF & $-5.76$ & $2.38$ & $0.52$ & $-7.45$ & $-10.31$ \\ & ooUCCSD & $-5.76$ & $2.38$ & $0.52$ & $-7.49$ & $-10.36$ \\ [1.0ex]	(10,24) & CCSD & $-5.79$ & $2.42$ & $0.48$ & $-5.75$ & $-8.64$ \\[1.0ex]
    \bottomrule \end{tabular}
\end{table}

\begin{table}[h!]
    \caption{One-bond nitrogen-hydrogen indirect nuclear spin-spin coupling constant $^1J_\mathrm{NH}$ in NH$_3$ for CASCI, UCCSD, CASSCF and ooUCCSD in one active space and full-space (10,24) CCSD with the 6-31G-J basis set.} \label{tab:nh3_nh}
    \centering
    \begin{tabular}{rlccccc}
        \toprule \multicolumn{2}{l}{Method} & DSO & PSO & SD & FC & Total \\ \midrule
(6,6) & CASCI & $0.14$ & $0.27$ & $0.06$ & $-1.87$ & $-1.40$ \\ & UCCSD & $0.14$ & $0.27$ & $0.06$ & $-1.86$ & $-1.39$ \\ & CASSCF & $0.14$ & $1.75$ & $0.11$ & $59.12$ & $61.12$ \\ & ooUCCSD & $0.14$ & $1.75$ & $0.11$ & $59.02$ & $61.02$ \\ [1.0ex] (10,24) & CCSD & $0.14$ & $1.73$ & $0.10$ & $55.89$ & $57.86$ \\[1.0ex]
    \bottomrule \end{tabular}
\end{table}

\begin{table}[h!]
    \caption{Two-bond hydrogen-hydrogen indirect nuclear spin-spin coupling constant $^2J_\mathrm{HH}$ in CH$_4$ for CASCI, UCCSD, CASSCF and ooUCCSD in one active space and full-space (10,28) CCSD with the 6-31G-J basis set.} \label{tab:ch4_hh}
    \centering
    \begin{tabular}{rlccccc}
        \toprule \multicolumn{2}{l}{Method} & DSO & PSO & SD & FC & Total \\ \midrule
(8,8) & CASCI & $-3.13$ & $0.46$ & $0.08$ & $-2.06$ & $-4.64$ \\ & UCCSD & $-3.13$ & $0.46$ & $0.08$ & $-2.07$ & $-4.65$ \\ & CASSCF & $-3.15$ & $1.14$ & $0.45$ & $-22.22$ & $-23.79$ \\ & ooUCCSD & $-3.15$ & $1.14$ & $0.45$ & $-22.25$ & $-23.81$ \\ [1.0ex]
(10,28) & CCSD & $-3.16$ & $1.16$ & $0.43$ & $-19.55$ & $-21.12$ \\[1.0ex]
    \bottomrule \end{tabular}
\end{table}

\begin{table}[h!]
    \caption{One-bond carbon-hydrogen indirect nuclear spin-spin coupling constant $^1J_\mathrm{CH}$ in CH$_4$ for CASCI, UCCSD, CASSCF and ooUCCSD in one active space and full-space (10,28) CCSD with the 6-31G-J basis set.} \label{tab:ch4_ch}
    \centering
    \begin{tabular}{rlccccc}
        \toprule \multicolumn{2}{l}{Method} & DSO & PSO & SD & FC & Total \\ \midrule
(8,8) & CASCI & $0.37$ & $-0.41$ & $0.20$ & $15.50$ & $15.65$ \\ & UCCSD & $0.37$ & $-0.41$ & $0.20$ & $15.54$ & $15.70$ \\ & CASSCF & $0.37$ & $1.40$ & $-0.02$ & $135.83$ & $137.58$ \\ & ooUCCSD & $0.37$ & $1.40$ & $-0.02$ & $135.29$ & $137.04$ \\ [1.0ex] (10,28) & CCSD & $0.37$ & $1.41$ & $0.05$ & $131.37$ & $133.20$ \\[1.0ex]

    \bottomrule \end{tabular}
\end{table}

\begin{table}[h!]
    \caption{One-bond carbon-oxygen indirect nuclear spin-spin coupling constant $^1J_\mathrm{CO}$ in CO for CASCI, UCCSD, CASSCF and ooUCCSD in five different active spaces and full-space (14,24) CCSD with the 6-31G-J basis set.} \label{tab:co_co}
    \centering
    \begin{tabular}{rlccccc}
        \toprule \multicolumn{2}{l}{Method} & DSO & PSO & SD & FC & Total \\ \midrule
(2,3) & CASCI & $0.01$ & $13.72$ & $3.75$ & $0.00$ & $17.48$ \\ & UCCSD & $0.01$ & $13.72$ & $3.75$ & $0.00$ & $17.48$ \\ & CASSCF & $0.01$ & $16.96$ & $-11.26$ & $13.11$ & $18.81$ \\ [1.0ex]	
(4,4) & CASCI & $0.01$ & $-15.26$ & $-8.12$ & $0.00$ & $-23.38$ \\ & CASSCF & $0.01$ & $20.65$ & $-3.92$ & $25.26$ & $42.00$ \\ & ooUCCSD & $0.01$ & $20.55$ & $-4.30$ & $25.06$ & $41.33$ \\ [1.0ex]	
(6,5) & CASCI & $0.01$ & $-2.07$ & $-4.51$ & $0.00$ & $-6.57$ \\ & UCCSD & $0.01$ & $-3.18$ & $-5.15$ & $0.00$ & $-8.32$ \\ & CASSCF & $0.01$ & $20.38$ & $-4.03$ & $22.12$ & $38.48$ \\ & ooUCCSD & $0.01$ & $20.31$ & $-4.39$ & $22.18$ & $38.11$ \\ [1.0ex]	
(6,6) & CASCI & $0.01$ & $-1.49$ & $-4.59$ & $1.69$ & $-4.38$ \\ & UCCSD & $0.01$ & $-2.58$ & $-5.24$ & $1.69$ & $-6.11$ \\ & CASSCF & $0.01$ & $17.84$ & $-4.51$ & $29.09$ & $42.42$ \\ & ooUCCSD & $0.01$ & $17.56$ & $-4.80$ & $28.67$ & $41.44$ \\ [1.0ex]	
(8,8) & CASCI & $0.01$ & $2.26$ & $-4.29$ & $4.59$ & $2.57$ \\ & UCCSD & $0.01$ & $1.31$ & $-4.90$ & $4.96$ & $1.38$ \\ & CASSCF & $0.01$ & $17.46$ & $-4.39$ & $30.60$ & $43.68$ \\ & ooUCCSD & $0.01$ & $17.41$ & $-4.50$ & $30.35$ & $43.26$ \\ [1.0ex] (14,24) & CCSD & $0.01$ & $16.57$ & $-5.06$ & $30.20$ & $41.72$ \\[1.0ex]
    \bottomrule \end{tabular}
\end{table}

\clearpage
\section{Breakdown of spin-spin coupling constants into individual contributions}

\begin{figure}[hbpt!]
    \centering
    \subfloat[DSO contribution]{\includegraphics[width=0.45\linewidth]{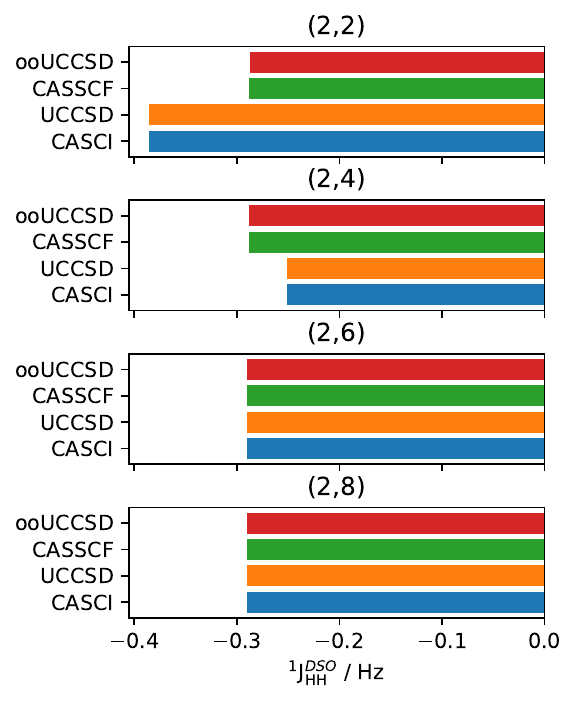}}
    \subfloat[SD contribution]{\includegraphics[width=0.45\linewidth]{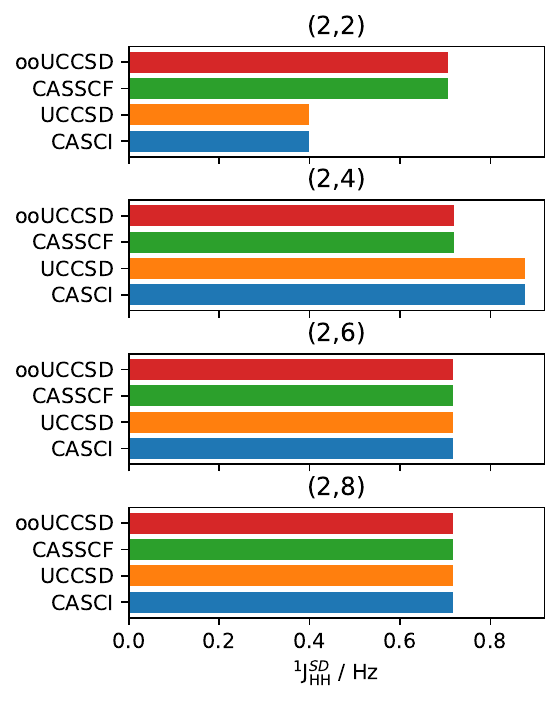}}
    \\ \subfloat[FC contribution]{\includegraphics[width=0.45\linewidth]{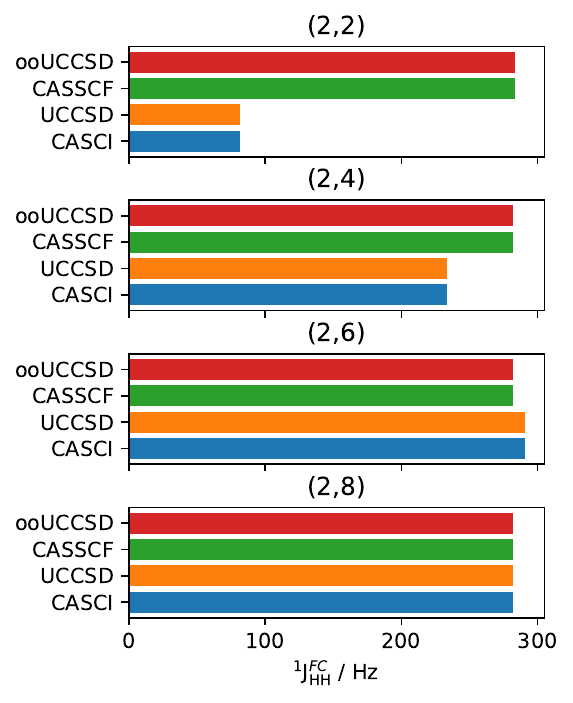}}
    \caption{Breakdown of spin-spin coupling constant of H$_2$ in a 6-31G-J basis into individual contributions. Note that the PSO contribution is zero for all methods and active spaces.}
    \label{fig:h2_hh_terms}
\end{figure}

\begin{figure}[h!]
    \centering
    \subfloat[DSO contribution]{\includegraphics[width=0.45\linewidth]{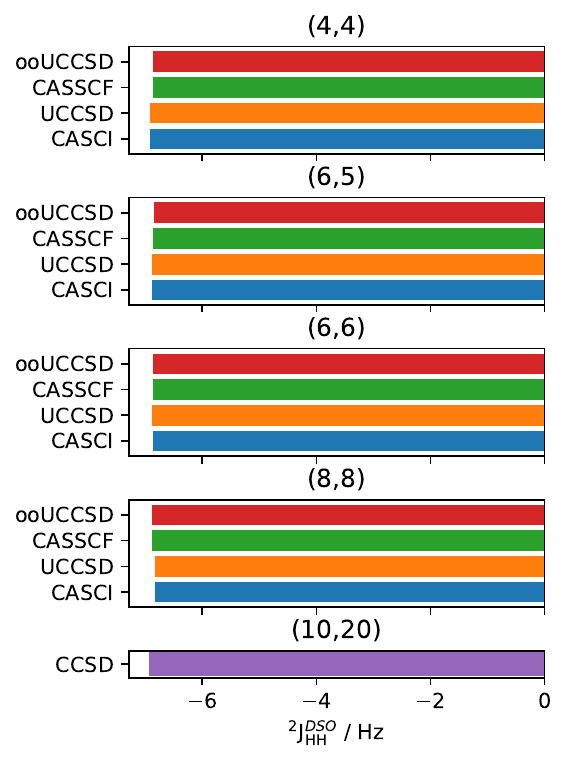}}
    \subfloat[PSO contribution]{\includegraphics[width=0.45\linewidth]{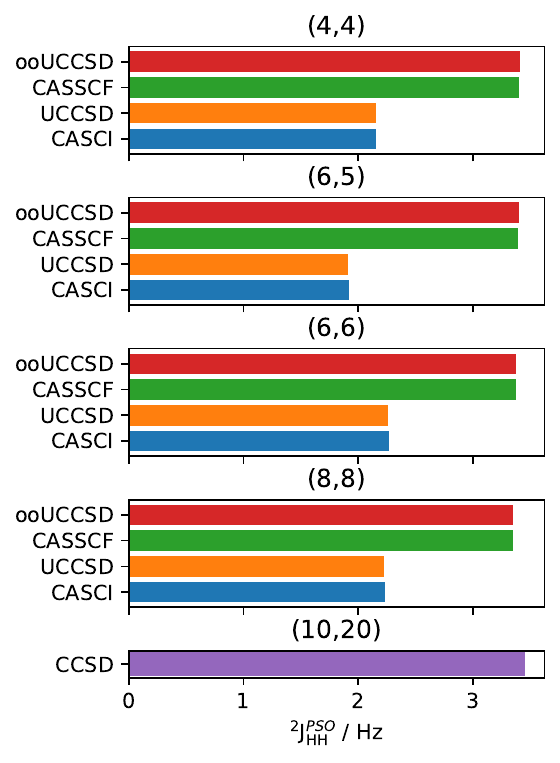}}
    \\ \subfloat[SD contribution]{\includegraphics[width=0.45\linewidth]{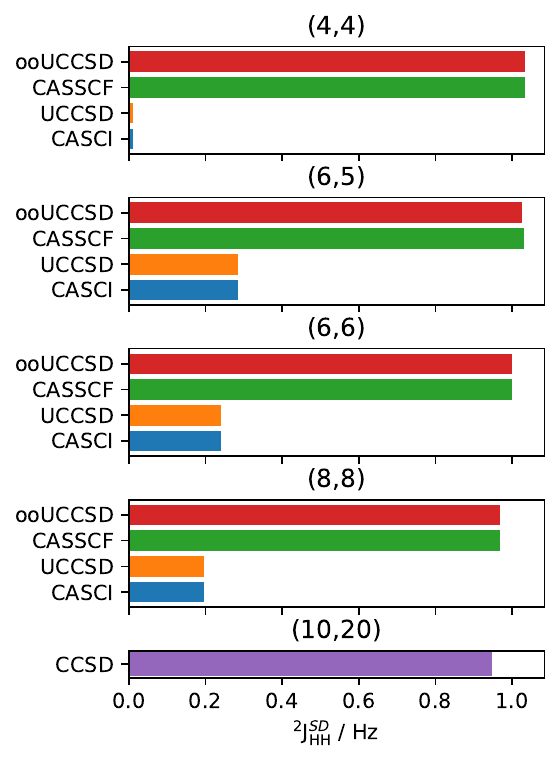}}
    \subfloat[FC contribution]{\includegraphics[width=0.45\linewidth]{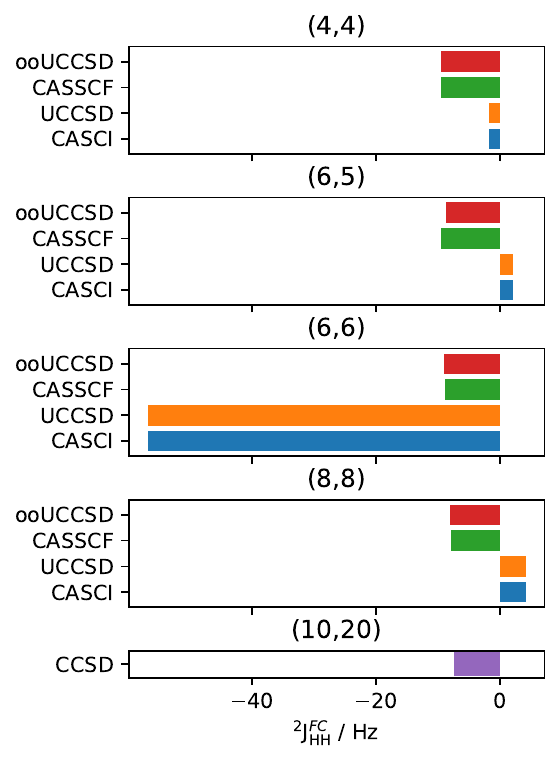}}
    \caption{Breakdown of the H-H spin-spin coupling constant of H$_2$O in a 6-31G-J basis into individual contributions.}
    \label{fig:oh2_hh_terms}
\end{figure}

\begin{figure}[h!]
    \centering
    \subfloat[DSO contribution]{\includegraphics[width=0.45\linewidth]{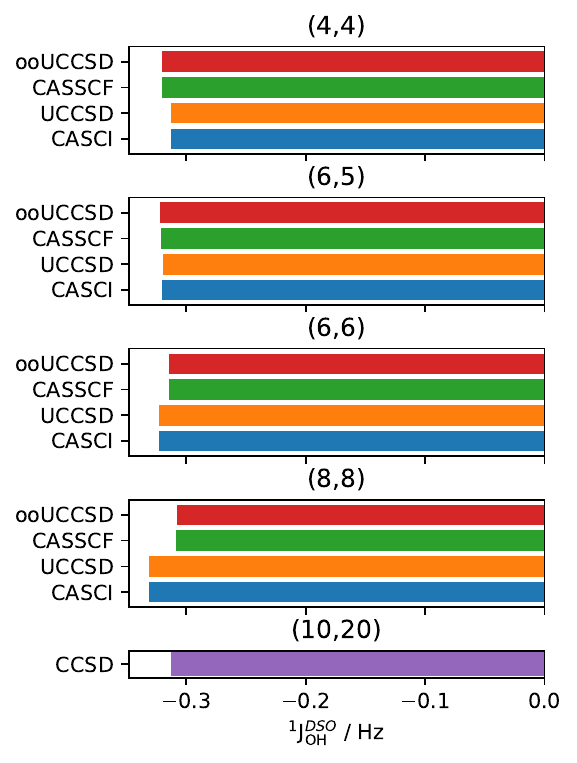}}
    \subfloat[PSO contribution]{\includegraphics[width=0.45\linewidth]{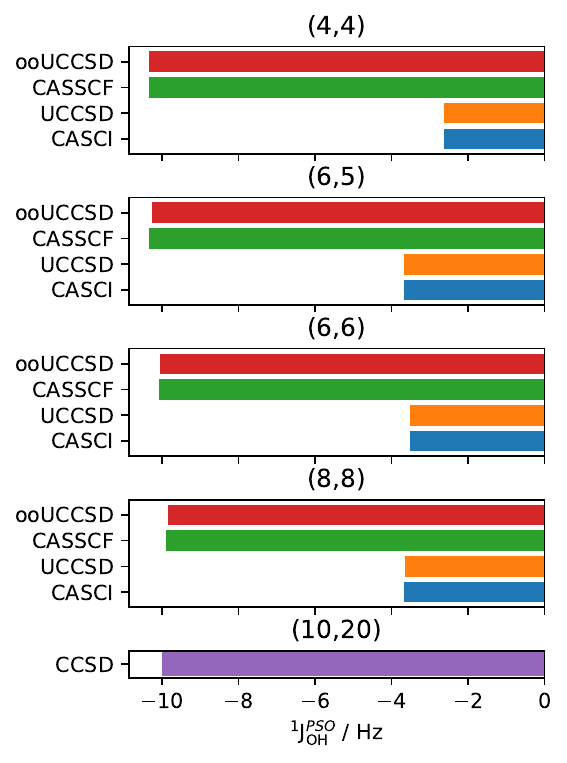}}
    \\ \subfloat[SD contribution]{\includegraphics[width=0.45\linewidth]{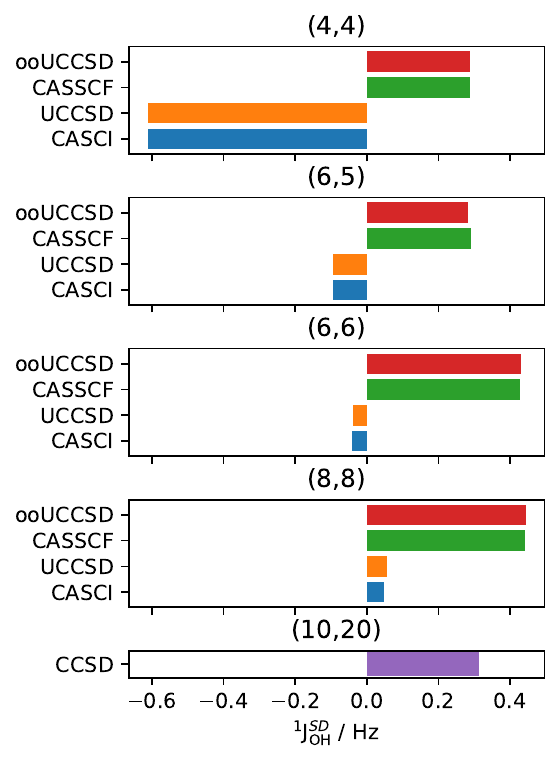}}
    \subfloat[FC contribution]{\includegraphics[width=0.45\linewidth]{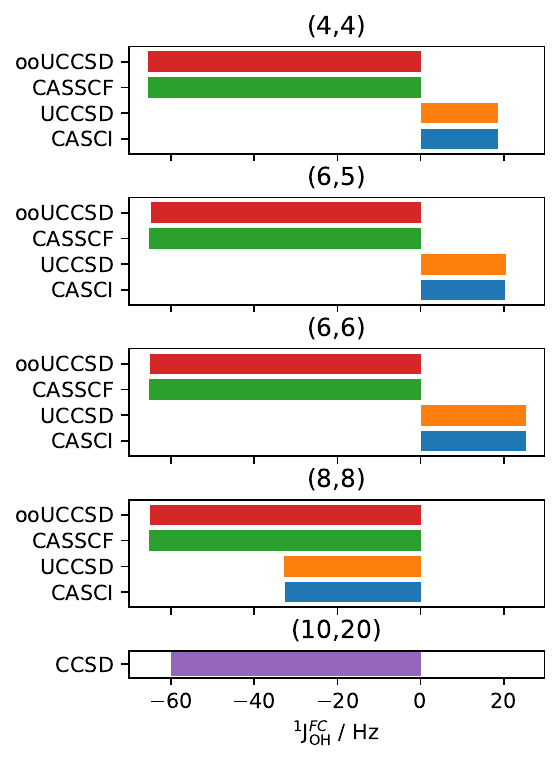}}
    \caption{Breakdown of the O-H spin-spin coupling constant of H$_2$O in a 6-31G-J basis into individual contributions.}
    \label{fig:oh2_oh_terms}
\end{figure}

\begin{figure}[h!]
    \centering
    \subfloat[DSO contribution]{\includegraphics[width=0.5\linewidth]{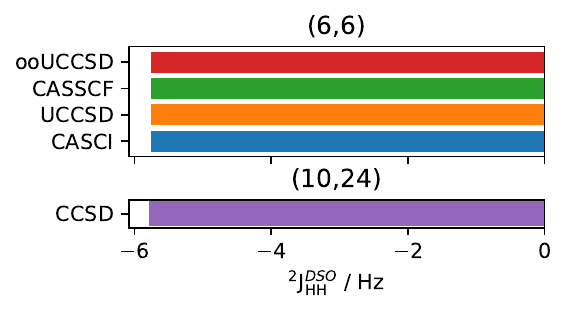}}
    \subfloat[PSO contribution]{\includegraphics[width=0.5\linewidth]{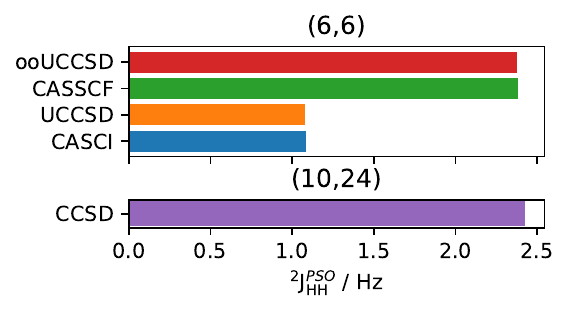}}
    \\ \subfloat[SD contribution]{\includegraphics[width=0.5\linewidth]{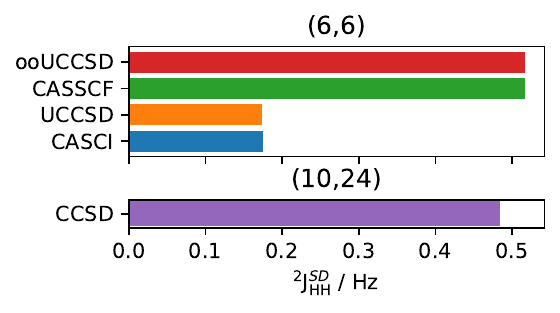}}
    \subfloat[FC contribution]{\includegraphics[width=0.5\linewidth]{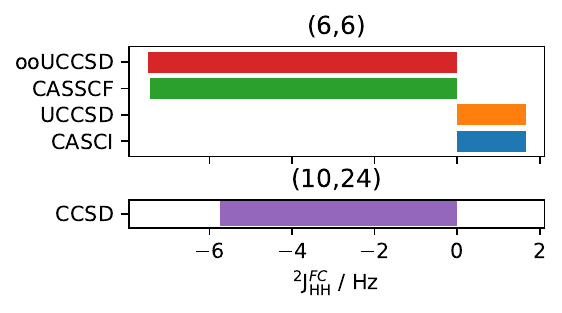}}
    \caption{Breakdown of the H-H spin-spin coupling constant of NH$_3$ in a 6-31G-J basis into individual contributions.}
    \label{fig:nh3_hh_terms}
\end{figure}

\begin{figure}[h!]
    \centering
    \subfloat[DSO contribution]{\includegraphics[width=0.5\linewidth]{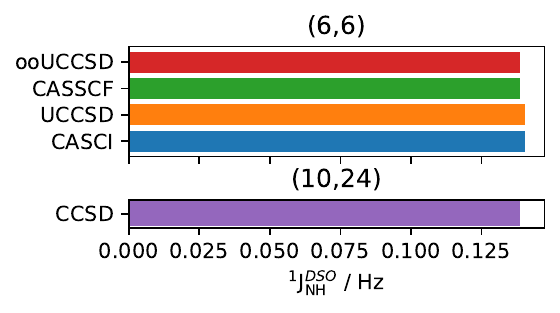}}
    \subfloat[PSO contribution]{\includegraphics[width=0.5\linewidth]{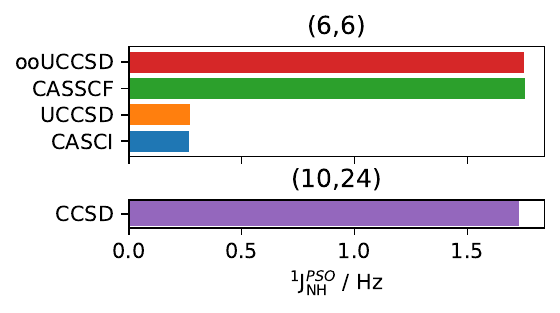}}
    \\ \subfloat[SD contribution]{\includegraphics[width=0.5\linewidth]{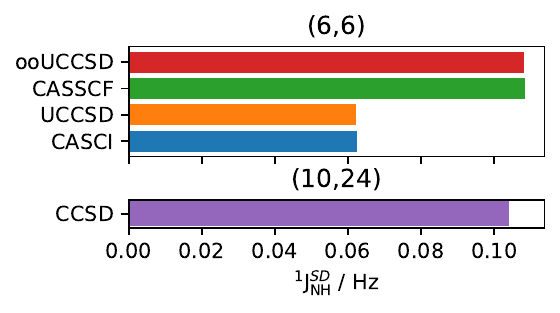}}
    \subfloat[FC contribution]{\includegraphics[width=0.5\linewidth]{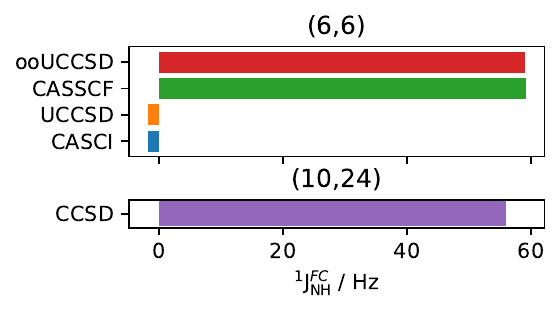}}
    \caption{Breakdown of the N-H spin-spin coupling constant of NH$_3$ in a 6-31G-J basis into individual contributions.}
    \label{fig:nh3_nh_terms}
\end{figure}

\begin{figure}[h!]
    \centering
    \subfloat[DSO contribution]{\includegraphics[width=0.5\linewidth]{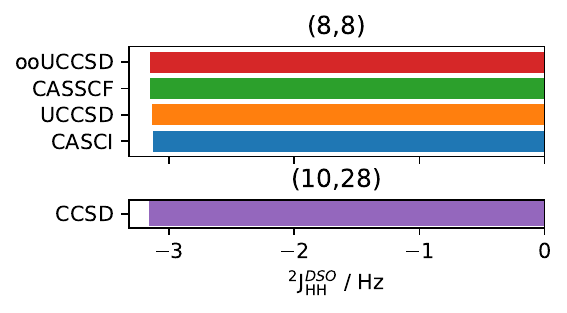}}
    \subfloat[PSO contribution]{\includegraphics[width=0.5\linewidth]{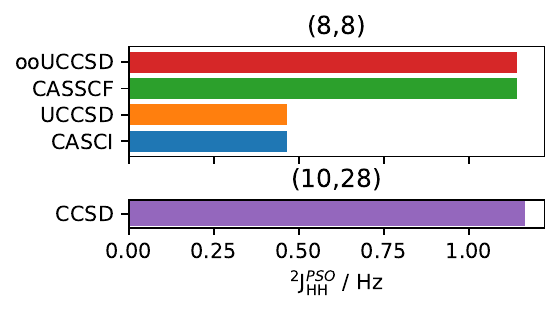}}
    \\ \subfloat[SD contribution]{\includegraphics[width=0.5\linewidth]{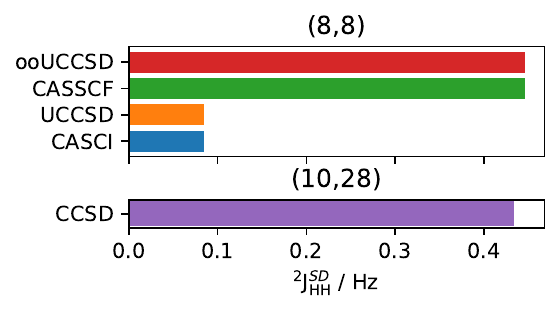}}
    \subfloat[FC contribution]{\includegraphics[width=0.5\linewidth]{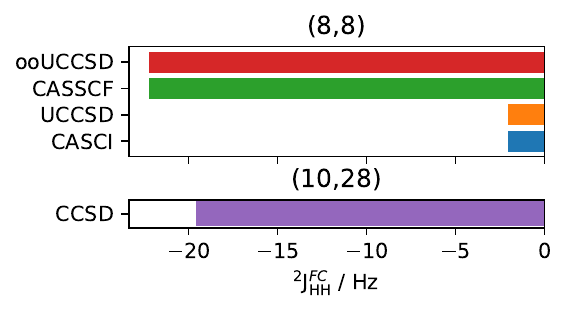}}
    \caption{Breakdown of the H-H spin-spin coupling constant of CH$_4$ in a 6-31G-J basis into individual contributions.}
    \label{fig:ch4_hh_terms}
\end{figure}

\begin{figure}[h!]
    \centering
    \subfloat[DSO contribution]{\includegraphics[width=0.5\linewidth]{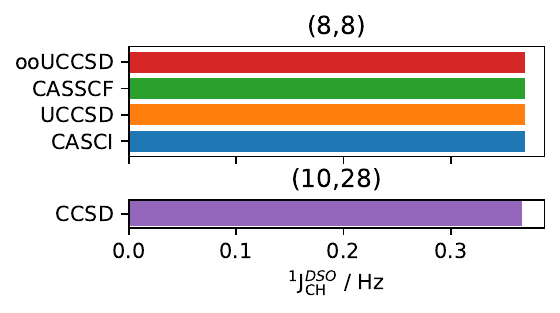}}
    \subfloat[PSO contribution]{\includegraphics[width=0.5\linewidth]{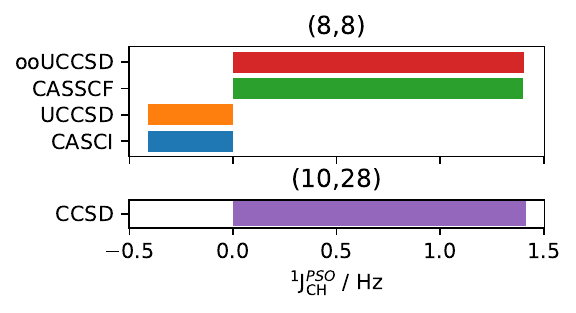}}
    \\ \subfloat[SD contribution]{\includegraphics[width=0.5\linewidth]{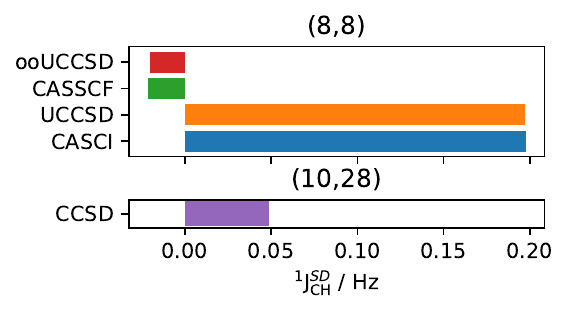}}
    \subfloat[FC contribution]{\includegraphics[width=0.5\linewidth]{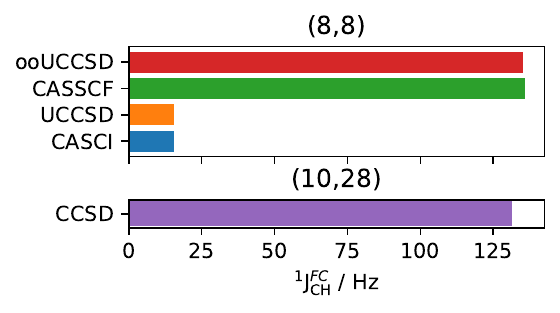}}
    \caption{Breakdown of the C-H spin-spin coupling constant of CH$_4$ in a 6-31G-J basis into individual contributions.}
    \label{fig:ch4_ch_terms}
\end{figure}

\begin{figure}[h!]
    \centering
    \subfloat[DSO contribution]{\includegraphics[width=0.5\linewidth]{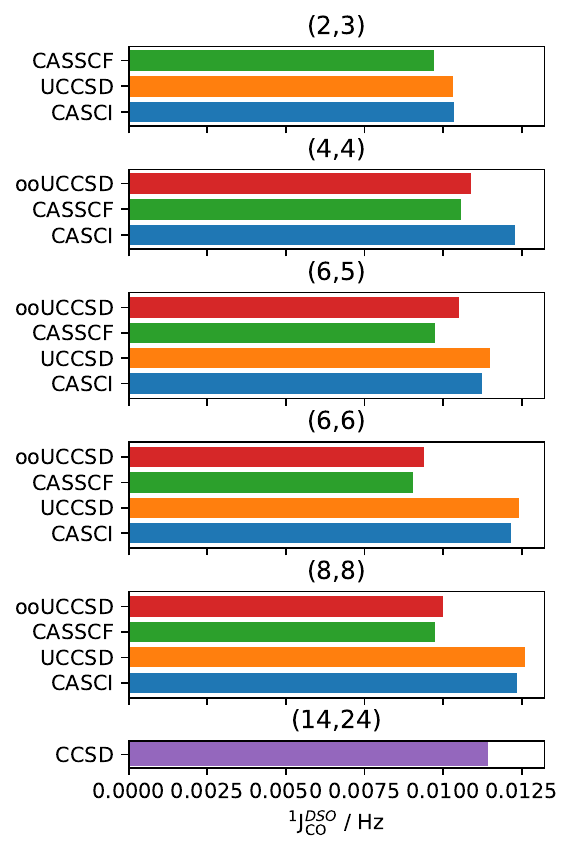}}
    \subfloat[PSO contribution]{\includegraphics[width=0.5\linewidth]{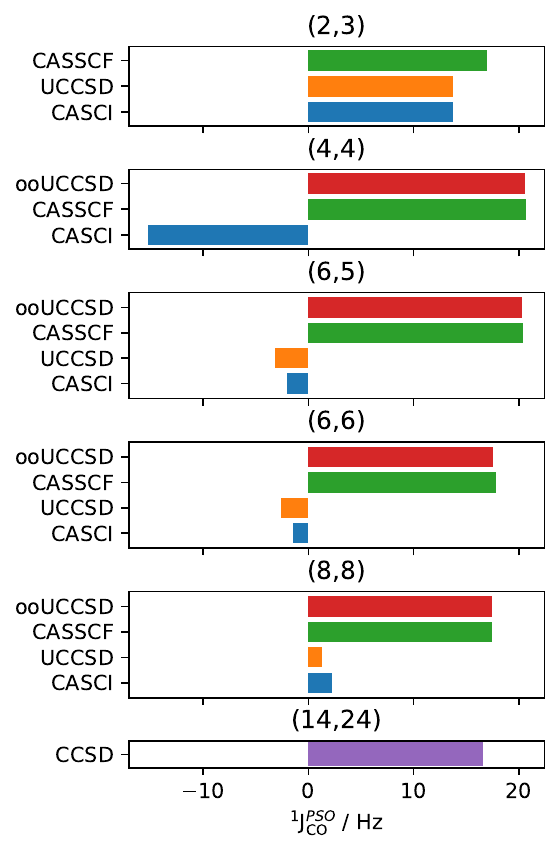}}
    \caption{Breakdown of the C-O spin-spin coupling constant of CO in a 6-31G-J basis into individual contributions.}
    \label{fig:co_co_terms}
\end{figure}
\begin{figure}
    \ContinuedFloat
    \subfloat[SD contribution]{\includegraphics[width=0.5\linewidth]{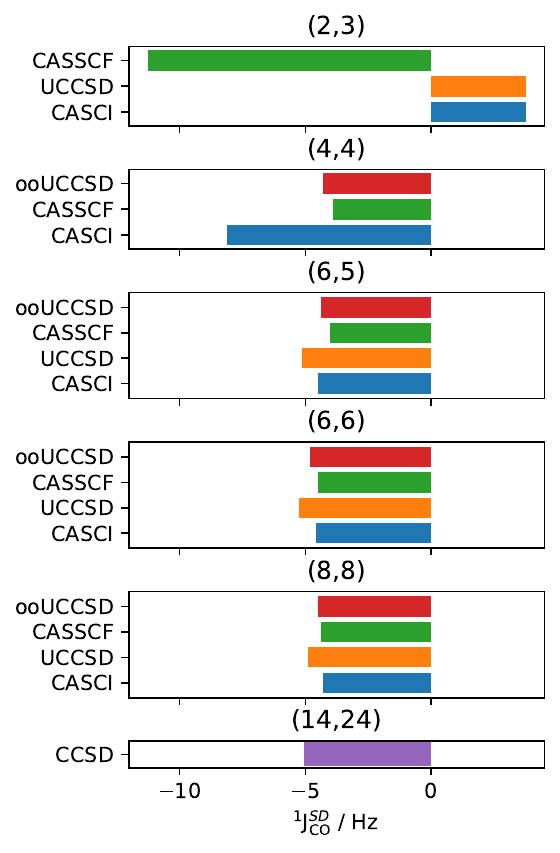}}
    \subfloat[FC contribution]{\includegraphics[width=0.5\linewidth]{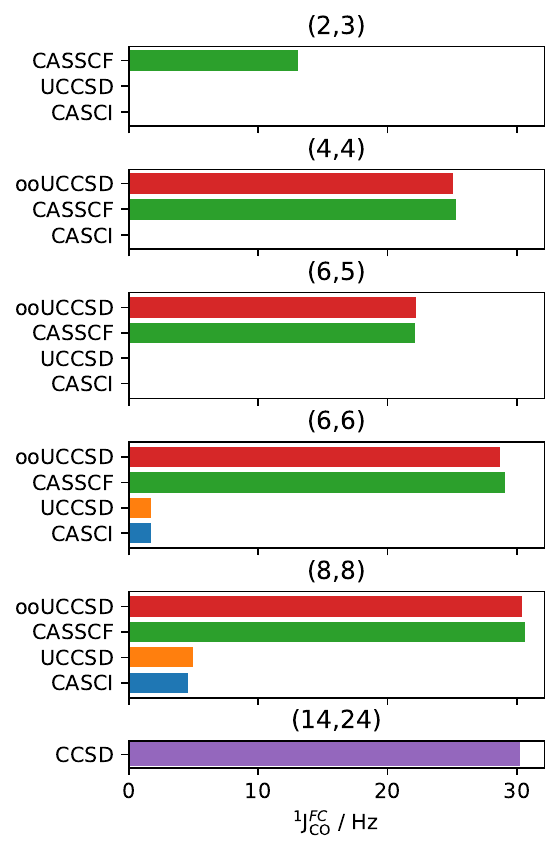}}
    \caption{Breakdown of the C-O spin-spin coupling constant of CO in a 6-31G-J basis into individual contributions. (cont.)}
\end{figure}

\end{document}